\documentclass[twocolumn]{aastex631}

\newcommand\kms{km$\,$s$^{-1}$}
\newcommand\Msol{M$_{\odot}$}

\newcommand{\hi}{H\,{\sc i}}
\newcommand{\hii}{H\,{\sc ii}}

\begin{document}

\title{Dark no more: The low luminosity stellar counterpart of a dark cloud in the Virgo cluster\footnote{Based on observations obtained with the Hobby-Eberly Telescope (HET), which is a joint project of the University of Texas at Austin, the Pennsylvania State University, Ludwig-Maximillians-Universitaet Muenchen, and Georg-August Universitaet Goettingen. The HET is named in honor of its principal benefactors, William P. Hobby and Robert E. Eberly.}}

\correspondingauthor{Michael G. Jones}
\email{jonesmg@arizona.edu}

\author[0000-0002-5434-4904]{Michael G. Jones}
\affiliation{Steward Observatory, University of Arizona, 933 North Cherry Avenue, Rm. N204, Tucson, AZ 85721-0065, USA}

\author[0000-0001-9165-8905]{Steven Janowiecki}
\affiliation{University of Texas, Hobby–Eberly Telescope, McDonald Observatory, TX 79734, USA}

\author[0009-0006-0732-3031]{Swapnaneel Dey}
\affiliation{Steward Observatory, University of Arizona, 933 North Cherry Avenue, Rm. N204, Tucson, AZ 85721-0065, USA}

\author[0000-0003-4102-380X]{David J. Sand}
\affiliation{Steward Observatory, University of Arizona, 933 North Cherry Avenue, Rm. N204, Tucson, AZ 85721-0065, USA}

\author[0000-0001-8354-7279]{Paul Bennet}
\affiliation{Space Telescope Science Institute, 3700 San Martin Drive, Baltimore, MD 21218, USA}

\author[0000-0002-1763-4128]{Denija Crnojevi\'{c}}
\affil{Department of Physics \& Astronomy, University of Tampa, 401 West Kennedy Boulevard, Tampa, FL 33606, USA}

\author[0000-0001-8245-779X]{Catherine E. Fielder}
\affiliation{Steward Observatory, University of Arizona, 933 North Cherry Avenue, Rm. N204, Tucson, AZ 85721-0065, USA}

\author[0000-0001-8855-3635]{Ananthan Karunakaran}
\affiliation{Department of Astronomy \& Astrophysics, University of Toronto, Toronto, ON M5S 3H4, Canada}
\affiliation{Dunlap Institute for Astronomy and Astrophysics, University of Toronto, Toronto ON, M5S 3H4, Canada}

\author[0000-0002-8990-1811]{Brian R. Kent}
\affiliation{National Radio Astronomy Observatory, 520 Edgemont Road, Charlottesville, VA 22903, USA}

\author[0009-0005-9612-4722]{Nicolas Mazziotti}
\affiliation{Steward Observatory, University of Arizona, 933 North Cherry Avenue, Rm. N204, Tucson, AZ 85721-0065, USA}

\author[0000-0001-9649-4815]{Bur\c{c}in Mutlu-Pakdil}
\affil{Department of Physics and Astronomy, Dartmouth College, Hanover, NH 03755, USA}

\author[0000-0002-0956-7949]{Kristine Spekkens}
\affiliation{Department of Physics and Space Science, Royal Military College of Canada P.O. Box 17000, Station Forces Kingston, ON K7K 7B4, Canada}
\affiliation{Department of Physics, Engineering Physics and Astronomy, Queen’s University, Kingston, ON K7L 3N6, Canada}



\begin{abstract}

We have discovered the stellar counterpart to the ALFALFA~Virgo~7 cloud complex, which has been thought to be optically dark and nearly star-free since its discovery in 2007. This $\sim$190~kpc long chain of enormous atomic gas clouds ($M_\mathrm{HI} \sim 10^9$~\Msol) is embedded in the hot intracluster medium of the Virgo galaxy cluster but is isolated from any galaxy. Its faint, blue stellar counterpart, BC6, was identified in a visual search of archival optical and UV imaging. Follow-up observations with the Green Bank Telescope, Hobby-Eberly Telescope, and Hubble Space Telescope demonstrate that this faint counterpart is at the same velocity as the atomic gas, is actively forming stars, and is metal-rich ($12 + \mathrm{(O/H)} = 8.58 \pm 0.25$). We estimate its stellar mass to be only $\log ( M_\ast/\mathrm{M_\odot}) \sim 4.4$, making it one of the most gas-rich stellar systems known. Aside from its extraordinary gas content, the properties of BC6 are entirely consistent with those of a recently identified class of young, low-mass, isolated, and star-forming clouds in Virgo, that appear to have formed via extreme ram pressure stripping events. We expand the existing discussion of the origin of this structure and suggest NGC~4522 as a likely candidate, however, the current evidence is not fully consistent with any of our proposed progenitor galaxies. We anticipate that other ``dark" gas clouds in Virgo may have similarly faint, star-forming counterparts. We aim to identify these through the help of a citizen science search of the entire cluster.

\end{abstract}

\keywords{Star formation regions (1565); Ram pressure stripped tails (2126); HI line emission (690); Galaxy interactions (600); Virgo Cluster (1772)}


\section{Introduction} \label{sec:intro}


Surveys of atomic gas (\hi) have a long history of identifying apparently dark structures made up of large quantities of neutral gas and (almost) no stars \citep[e.g.][]{Giovanelli+1989,Chengalur+1995,Verdes-Montenegro+2001,Davies+2004,Minchin+2005,Haynes+2007,Kent+2007,Taylor+2012,Wong+2021,Jozsa+2022,Jones+2023}. \hi \ is prone to forming such structures as it is typically the most loosely bound baryonic component of a galaxy's disk and is thus the most easily removed by external forces. Furthermore, as the distribution of \hi \ often extends well beyond the stellar disk of a galaxy it is relatively straightforward to strip gas, but no stars, thereby producing optically dark clouds.

Although some of the structures identified in the references above are still connected to their parent galaxies, some are detached, isolated, and apparently dark \hi \ clouds with no clear point of origin. Such objects span several orders of magnitude in \hi \ mass and are certainly not a single type of object. At high-mass and intermediate masses \citet{Kent+2007} and \citet{Taylor+2012} identified several dark \hi \ clouds ($7 \lesssim \log M_\mathrm{HI}/\mathrm{M_\odot} \lesssim 9$) in the direction of the Virgo cluster and the Five-hundred-meter Aperture Spherical Telescope and the Green Bank Telescope (GBT) have recently identified lone \hi \ clouds that may be ancient objects \citep{Zhou+2023,Karunakaran+2024,O'Neil+2024}. At the low-mass end, \citet{Adams+2013} identified a sample of optically dark, compact \hi \ clouds, some of which could be very low mass halos in or near the Local Group. 

When dealing with \hi-only detections of local objects (e.g. $cz \lesssim 5000$~\kms) it is challenging to determine accurate distances, and these systems may not be as they first appear. \citet{Adams+2015}, \citet{Bellazzini+2015}, and \citet{Sand+2015} all, almost simultaneously, realized that one of the \citet{Adams+2013} objects (AGC~226067) did contain a small number of stars and was $\sim$10$\times$ farther away and $\sim$100$\times$ more massive than first thought. This object, now commonly referred to as SECCO~1, was found to be an isolated pocket of recent star formation (SF) embedded in the hot intra-cluster medium (ICM) of the Virgo cluster. Its rich metallicity \citep{Beccari+2017} and exclusively young stellar population \citep{Sand+2017} indicated that it must have formed recently from stripped gas, but there was no candidate parent galaxy within over 200~kpc. 

\citet{Sand+2015} searched for other objects with similar optical and UV morphology to SECCO~1, identifying five additional candidates in Virgo. With Hubble Space Telescope (HST) imaging, plus Jansky Very Large Array (VLA), GBT, and Multi Unit Spectroscopic Explorer (MUSE) observations \citep{Jones+2022,Jones+2022b,Bellazzini+2022} we demonstrated that four\footnote{The fifth object is resolved as a background galaxy group in HST imaging \citep{Jones+2022b}.} of these are analogous to SECCO~1 (though not all are \hi-rich) and argued that they represent a new class of object, which we will refer to simply as ``blue blobs." The only viable mechanism to explain both the isolation and young ages of blue blobs is high speed ram pressure stripping of \hi-bearing galaxies falling into the cluster \citep{Jones+2022b}. Even so, it is remarkable that they survived traversing 100s of kpc through the hot ICM, and may be the first examples of isolated star-forming clouds supported (in part) by external pressure \citep{Burkhart+2016,Bellazzini+2018,Calura+2020,Jones+2022b}.

With the goal of better understanding the population statistics of these new objects we embarked on a uniform search for candidates covering the entire cluster (described in \S\ref{sec:search}). When we began following up our initial candidates (\S\ref{sec:obs}), one (known as BC6) was found to be coincident (both in position and velocity) with the largest of the ``dark'' \hi \ cloud complexes identified by \citet{Kent+2007}, ALFALFA~Virgo~7. In a follow-up investigation, \citet{Kent+2009} found that VCC~1357 overlaps with this \hi \ structure, but was deemed to likely be in projection and not physically associated with the gas due to its color and dwarf spheroidal morphology. Another faint coincident object was noted by \citet{Kent+2009}, but more recent imaging clearly indicates that this is a distant background group of galaxies.
Thus, no convincing stellar counterpart had been identified for this \hi \ structure since its discovery. Here we show (\S\ref{sec:results}) that ALFALFA~Virgo~7 has a definite stellar counterpart that is over 20,000 times less massive than the \hi, and is made up exclusively of stars that recently formed in situ. The stellar properties of this object closely match those of the recently identified blue blobs \citep{Jones+2022b}. BC6 differs only in its exceptional \hi \ content. In \S\ref{sec:discuss} we discuss the potential point of origin of this object, and its likely fate. We present our conclusions in \S\ref{sec:conclusions}.

Throughout this letter we attempt to use ``BC6" to refer to the stellar component which we have identified, and the ``\hi \ cloud complex'' to refer to the (previously dark) \hi \ structure identified by \citet{Kent+2007}. We also assume throughout that BC6 and the associated \hi \ complex are in the Virgo cluster at a distance of 16.5~Mpc \citep{Mei+2007}. We note that this assumption is consistent with all observations, in particular the color--magnitude diagram (\S\ref{sec:HST_data} \& \S\ref{sec:stellar_pop}).

\section{Identification of candidates} \label{sec:search}

\begin{figure*}
    \centering
    \includegraphics[width=\columnwidth]{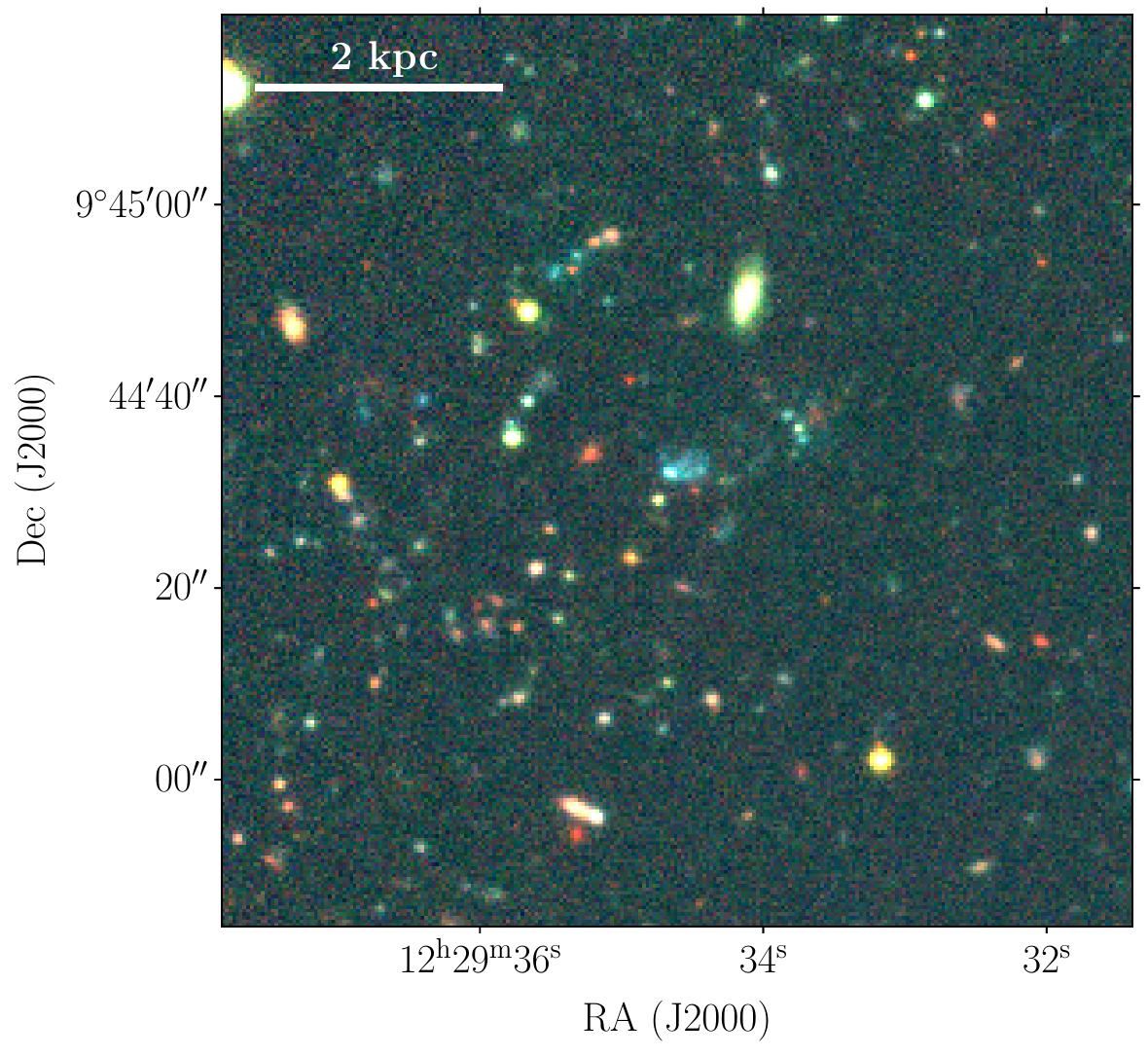}
    \includegraphics[width=\columnwidth]{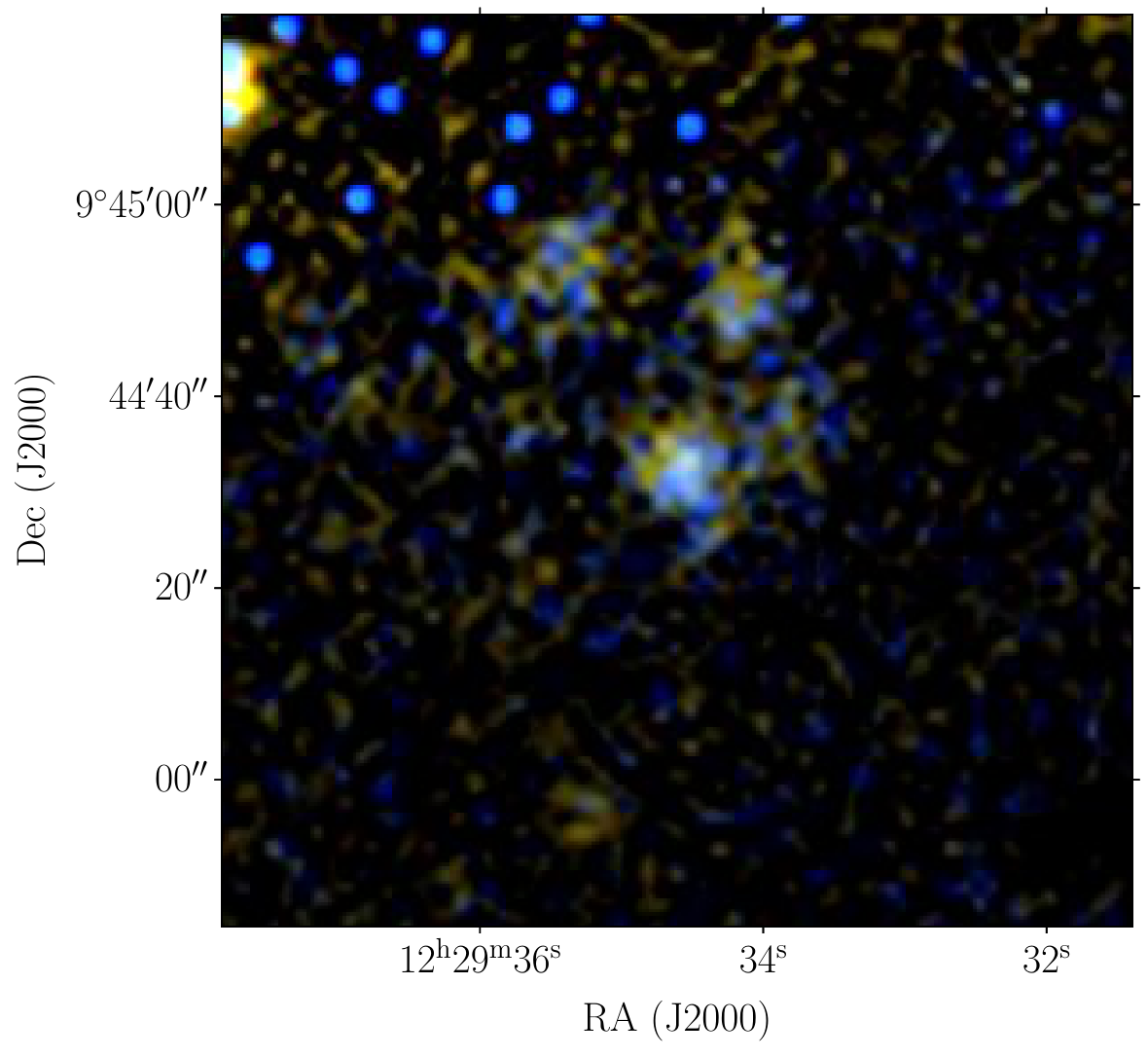}
    \caption{\textit{Left}: NGVS $ugi$ image of BC6, similar to the image in which it was originally identified. \textit{Right}: GALEX NUV+FUV image similar to that used for the original identification and covering the same field as in the left panel. The source is deliberately unlabeled, but its components are circled in Figure~\ref{fig:HI_map} (right), which shows a similar FoV. Without the comparison of the optical and UV images blue blobs would be almost impossible to identify. The bright blue speckles in the top-left of the right panel are an artifact from the GALEX FUV data.}
    \label{fig:BC6_NGVS}
\end{figure*}

The blue blob candidates presented in \citet{Jones+2022b} arose from a cursory search \citep{Sand+2015} of the Virgo cluster and there are likely additional examples. In an attempt to identify additional objects we began a visual search of the available imaging within the cluster from the Next Generation Virgo cluster Survey \citep[NGVS,][]{Ferrarese+2012}, the Dark Energy Camera Legacy Survey \citep[DECaLS][]{Dey+2019}, and the Galaxy Evolution Explorer \citep[GALEX][]{Martin+2005}. A full description of this search will be presented in Dey et al. (in prep.); here we provide a brief summary.

Matching 3\arcmin$\times\,$3\arcmin \ image cutouts from the three surveys were produced, covering the full area of the NGVS. The DECaLS $gri$ color images were taken directly from the Legacy Surveys Viewer\footnote{\url{www.legacysurvey.org/viewer}}, as were the GALEX NUV+FUV images. In the case of NGVS, we produced our own RGB images from the $u$, $g$ and $i$ filters, with a stretch designed to bring out faint, blue features (saturating most normal galaxies in the cluster). These image cutouts were uploaded to the \texttt{Zooniverse} website where team members (primarily M.~Jones, S.~Dey, and D.~Sand) searched for objects that had UV emission (or were exceptionally blue in the optical), but had extremely faint, irregular, and clumpy appearances in the optical images.\footnote{This initial search has since evolved into a full citizen science project that will be presented in detail in Mazziotti et al. (in prep) and Dey et al. (in prep). It is available at \url{https://www.zooniverse.org/projects/mike-dot-jones-dot-astro/blobs-and-blurs-extreme-galaxies-in-clusters}.} 

The most compelling blue blob candidates to emerge from this search, one of which was Blue Candidate 6 (BC6, Figure~\ref{fig:BC6_NGVS}), were selected for follow-up observations with GBT, HST, and HET, to search for \hi \ gas content, resolve their stellar populations, and search for optical emission lines, respectively. These observations are discussed in \S\ref{sec:obs}.


\section{Observations} \label{sec:obs}

BC6 lies in the direction of the Virgo cluster approximately 2.7$^\circ$ south of M~87 and 1.7$^\circ$ north of M~49. That is, roughly halfway been the Virgo A and B clouds \citep{Binggeli+1987}. Typical cluster members in this direction have line of sight velocities in the range $850 \lesssim cz/\mathrm{km\,s^{-1}} \lesssim 1200$, but the scatter is several hundred \kms \ beyond either side of this range \citep{Mei+2007}. There is a bridge of diffuse X-ray emission that connects M~87 to M~49 \citep{Bohringer+1994}, indicating that the hot ICM is higher density in this region than would be expected given its separation from the cluster core. BC6 lies roughly on this bridge.

\subsection{NGVS, DECaLS, and GALEX}

BC6 appears to have three main components (Figure~\ref{fig:BC6_NGVS}). The largest and brightest is just south of a background galaxy that is visible in both optical and UV images. This main component (BC6a) contains a bright knot of FUV  emission (blue in Figure~\ref{fig:BC6_NGVS}, right) and appears as a faint, clumpy, blue smudge in the NGVS image. The other two components are even fainter and are to the NE and east of the brightest component. BC6b (to the NE) has faint, but still clearly discernible UV emission, while BC6c (to the east) is barely detected in UV and appears only as a few blue points in the NGVS image.

Aperture photometry was performed on the NGVS $g$ and $i$ filters and GALEX NUV using the Aperture Photometry Tool \citep{Laher+2012}. Due to the irregular morphology of BC6, elliptical apertures were manually tailored for each component and the median sky level subtracted based on annuli beyond them. Galactic extinction was corrected using the dust maps of \citet{Schlegel+1998} and the filter coefficients from \citet{Schlafly+2011} and \citet{Wyder+2007}. Using standard conversions \citep{Iglesias-Paramo+2006,Morrissey+2007} we converted the GALEX NUV flux into a star formation rate (SFR) estimate. These photometry values are listed in Table~\ref{tab:props}.

\subsection{Green Bank Telescope}

After the identification of BC6 and several other candidates in the NGVS and GALEX imaging, we rapidly embarked on a follow-up program with a joint GBT-HST project, GBT-22B-080 (PI: M.~Jones), to search for \hi \ gas in all candidates and to obtain high resolution imaging of the stellar population of BC6. In addition, we followed up candidates with a Hobby-Eberly Telescope (HET) project (PI: S.~Janowiecki) to obtain optical spectroscopy of nebular emission lines (\S\ref{sec:HET_data}).

\begin{figure}
    \centering
    \includegraphics[width=\columnwidth]{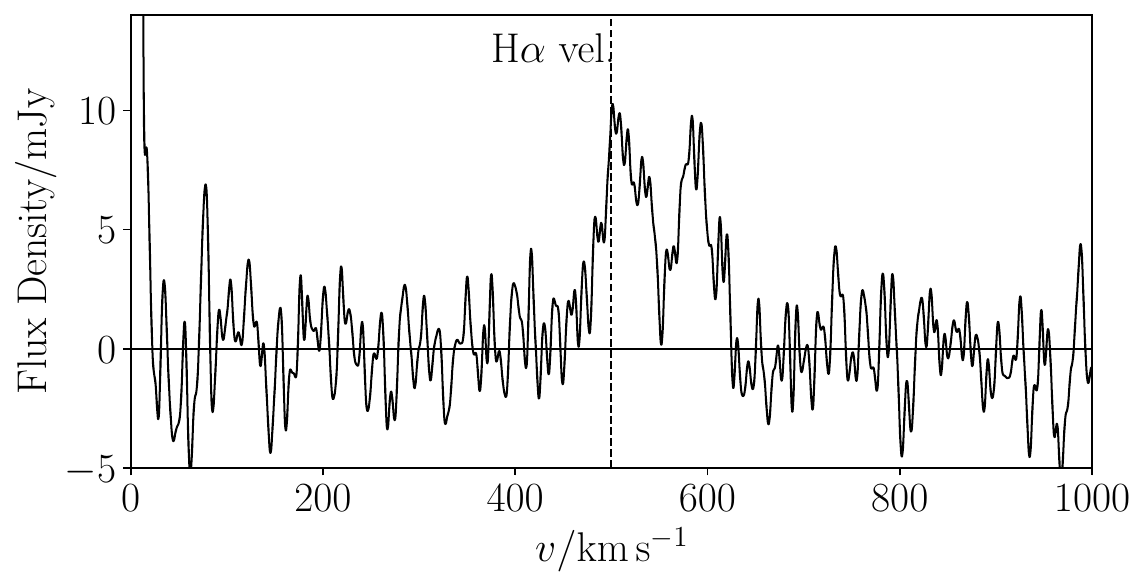}
    \caption{GBT \hi \ spectrum at the location of BC6. The vertical dashed line indicates the velocity of BC6 determined from H$\alpha$ line emission (\S\ref{sec:HET_data}). The profile is double horned because the large GBT beam includes some emission from the adjacent \hi \ cloud just to the SE \citep[denoted as C4 by][]{Kent+2009}, which is at a slightly higher velocity.}
    \label{fig:GBT_spec}
\end{figure}

Before it was realized that BC6 was the stellar counterpart to the \hi \ structure identified by \citet{Kent+2007}, and that it therefore already had \hi \ observations, it was included as a target in the GBT project. BC6 was observed in April 2023 with six 10~min on/off integrations with the VEGAS (Versatile GBT Astronomical Spectrometer) backend. The data were reduced using standard procedures in \texttt{GBTIDL}.\footnote{https://gbtidl.nrao.edu} After smoothing to a spectral resolution of 5~\kms, the resulting spectrum (Figure~\ref{fig:GBT_spec}) has an rms noise of 1.77~mJy. \hi \ line emission is strongly detected in the 9.1\arcmin\ GBT beam along the BC6 line-of-sight between 475~\kms\ -- 625~\kms.

\subsection{ALFALFA}

\begin{figure*}
    \centering
    \includegraphics[width=0.955\columnwidth]{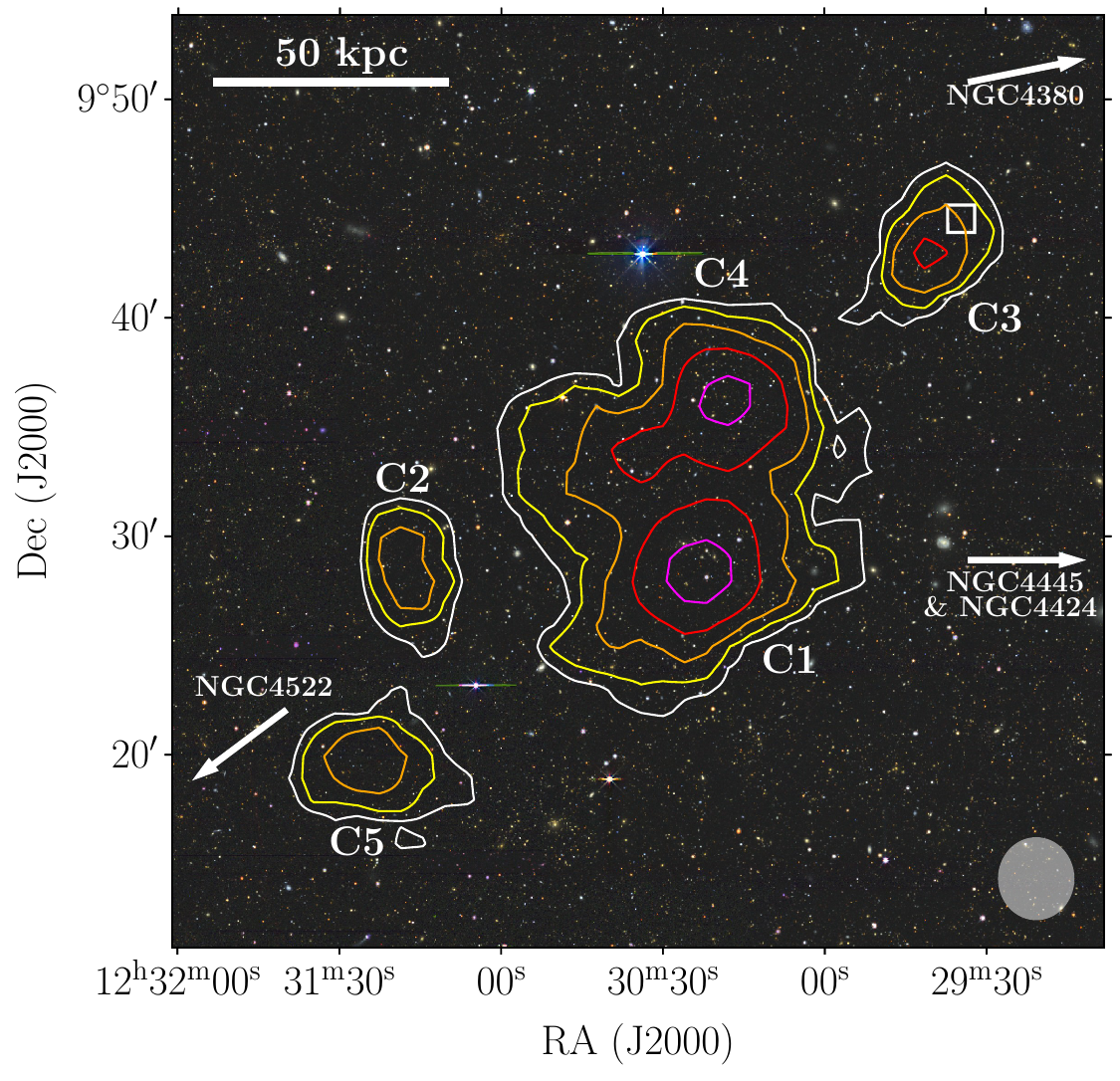}
    \includegraphics[width=\columnwidth]{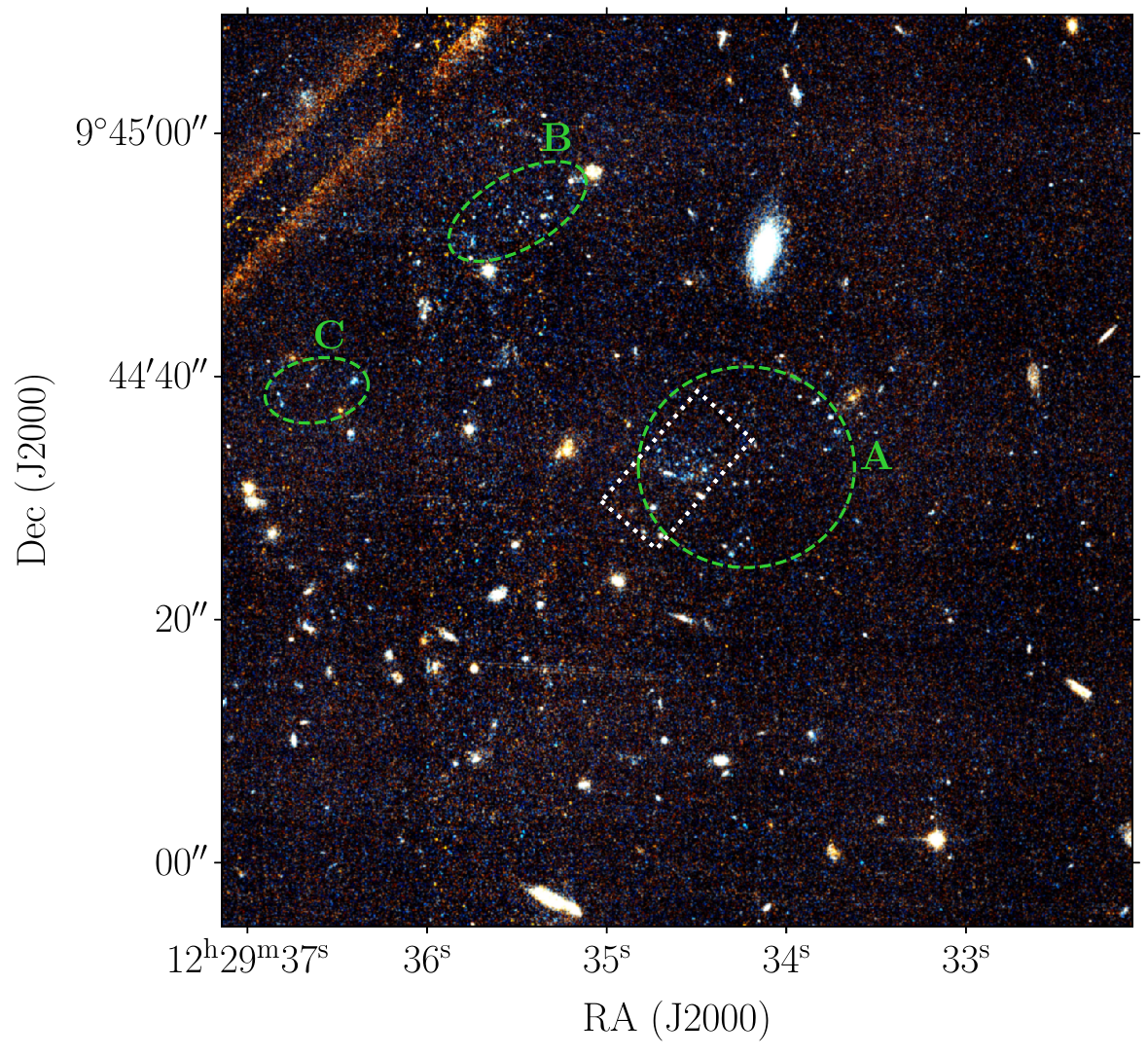}
    \caption{\textit{Left}: ALFALFA \hi \ contours overlaid on a $griz$ DECaLS image. BC6 is coincident with a cloud at the NW tip of the \hi \ complex (and is at the same velocity, cf. Figure~\ref{fig:GBT_spec}). The five component clouds of the \hi \ complex are labeled C1-5 as in \citet{Kent+2009}. The white box (inside cloud C3) indicates the location of BC6 and shows the size of the FoV of the HST image (right panel). The lowest \hi \ column density contour level is $2.2\times10^{18}\;\mathrm{cm^{-2}}$ (over 20~\kms) and each subsequent contour is double the previous. The ALFA beam size is shown in the bottom right corner. The white arrows around the edge of the panel indicate the directions toward some of the neighboring galaxies discussed in \S\ref{sec:origin} and Table~\ref{tab:progen}. \textit{Right}: HST ACS F814W+F606W color image of BC6. The three components are circled with dashed green ellipses. These are the regions used to produce the CMD of BC6 (Figure~\ref{fig:CMD}). The white dotted rectangle indicates the region targeted with HET/LRS2-B.}
    \label{fig:HI_map}
\end{figure*}

The Arecibo Legacy Fast ALFA (Arecibo L-band Feed Array), or ALFALFA, survey \citep{Giovanelli+2005,Haynes+2011,Haynes+2018} mapped approximately 3000~sq~deg of the sky for \hi \ line emission, including the Virgo cluster. 
\citet{Kent+2007} originally identified the \hi \ cloud complex in a preliminary ALFALFA data set. These data have a channel width of $\sim$5~\kms, a pixel size of 1\arcmin, and a spatial resolution of approximately 3.8\arcmin. We created our own version of the moment zero map of the \hi \ emission in the cloud complex using the \texttt{SoFiA} \citep{SoFiA,Serra+2015} masking tool, adopting the standard smooth and clip algorithm with spatial smoothing kernels of 4.0 and 8.0\arcmin, with both no spectral smoothing and smoothing over two and four channels ($\sim$10 and $\sim$20~\kms). We set a detection threshold of 3.5$\sigma$ in order to include relatively faint emission, but coupled this with a high reliability threshold (95\%) in order to ensure that only sources that are high S/N overall were retained. The moment zero map of the emission included in the \texttt{SoFiA} mask is shown in Figure~\ref{fig:HI_map} (left) as contours on a DECaLS optical image.

We estimate the \hi \ mass of the entire structure by integrating all of the flux within the \texttt{SoFiA} mask and assuming a distance of 16.5~Mpc. This gives a value of $\log (M_\mathrm{HI}/\mathrm{M_\odot}) = 8.83 \pm 0.04$. We note that this is about 0.1~dex higher than the value in \citet{Kent+2009}. However, in that work they considered each cloud component as a separate source, which may have led to some extended flux being missed. Whereas \texttt{SoFiA}'s spatial and spectral smoothing improves our ability to recover extended emission. In addition, \citet{Minchin+2019} used deeper Arecibo observations to estimate a total \hi \ mass of $\log (M_\mathrm{HI}/\mathrm{M_\odot}) = 9.1$, arguing that the additional flux (compared to ALFALFA) comes from low column density, extended emission.\footnote{\citet{Minchin+2019} also applied a 33\% (0.1~dex) correction factor to account for the Arecibo/ALFA sidelobe response. Both \citet{Kent+2009} and \citet{Minchin+2019} adopt a distance of 16.7~Mpc for Virgo, which would increase their mass estimates by 2.4\% (0.01~dex) relative to ours.}

\subsection{Hubble Space Telescope imaging} \label{sec:HST_data}

\begin{figure*}
    \centering
    \includegraphics[width=0.5\columnwidth]{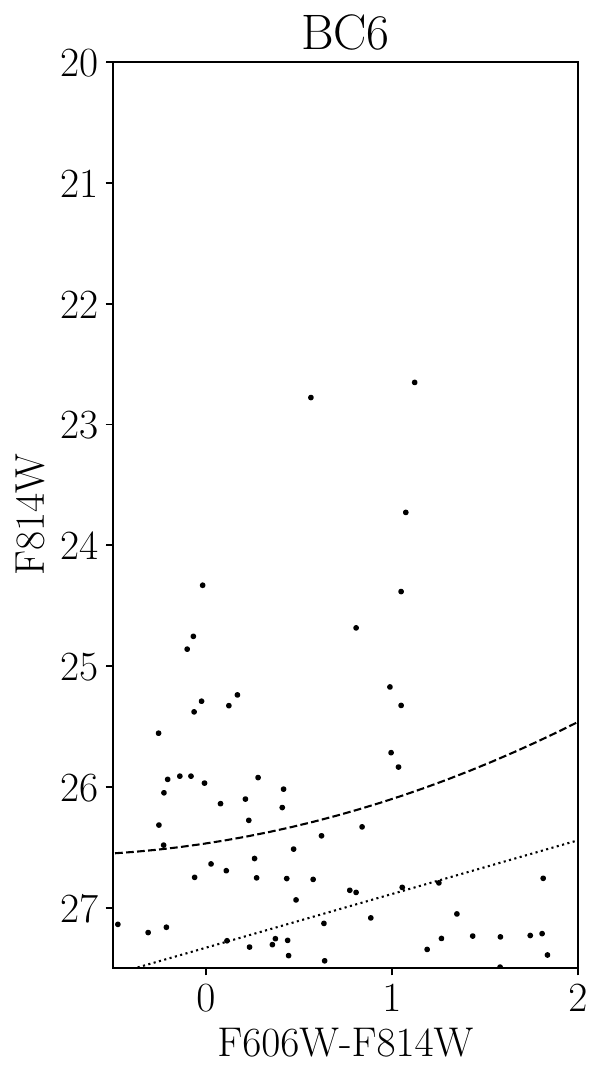}
    \includegraphics[width=0.5\columnwidth]{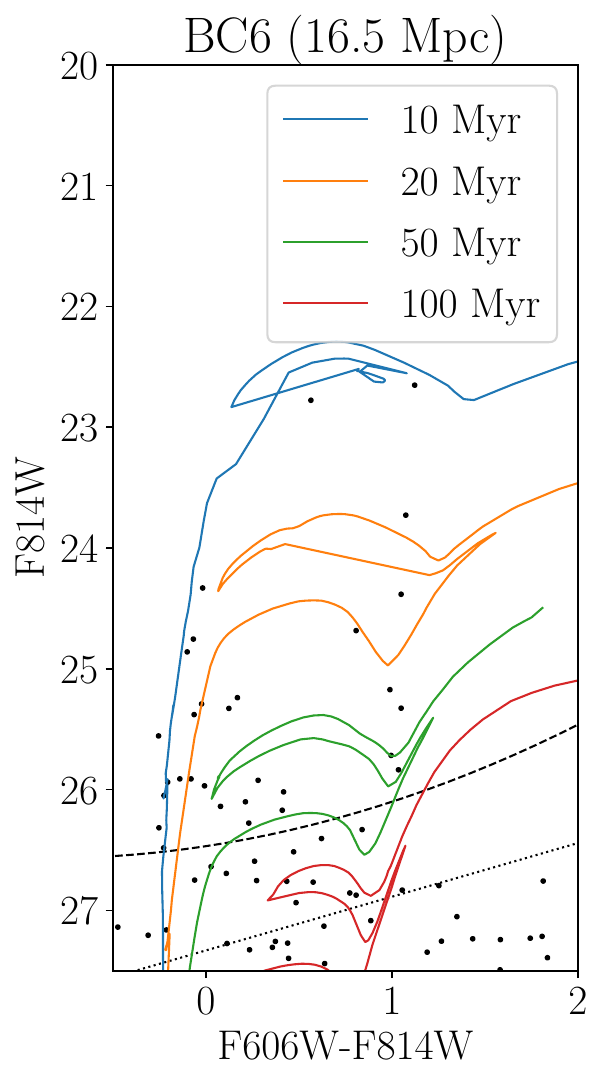}
    \includegraphics[width=0.5\columnwidth]{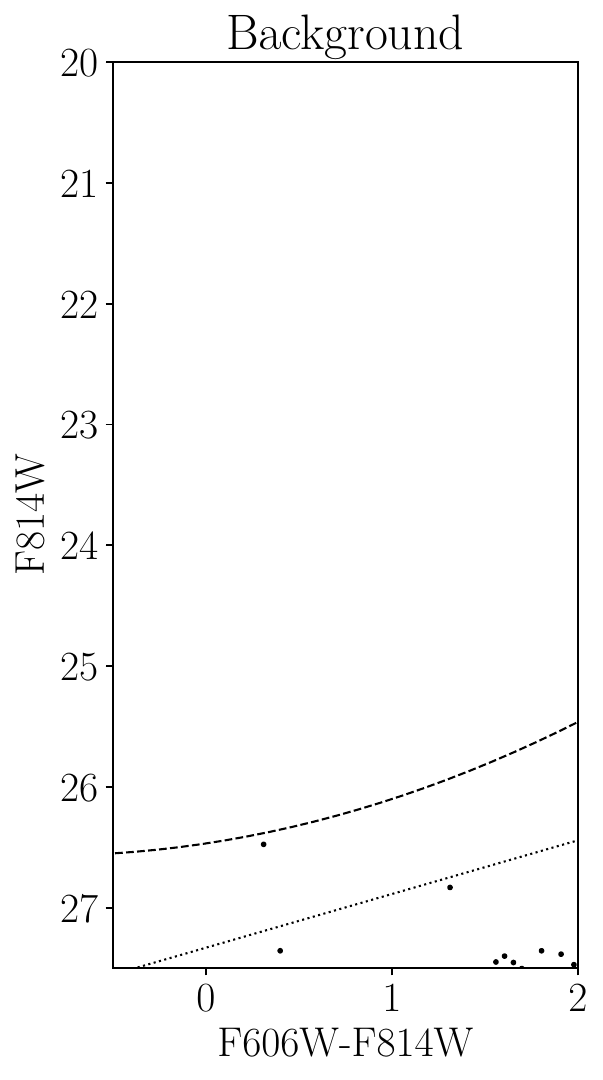}
    \caption{\textit{Left}: CMD of BC6 within the regions shown in Figure~\ref{fig:HI_map} (right). \textit{Center}: The same CMD with PARSEC isochrones for stellar populations with ages between 10 and 100~Myr. All isochrones have a metallicity of $[\mathrm{M/H}]=-0.1$, roughly matching that measured from emission lines (Table~\ref{tab:lines}), and are scaled to an assumed distance of 16.5~Mpc. \textit{Right}: CMD of a blank sky region (equal in area to the sum of the apertures in the left panel) of the HST image away the target. In all panels the dashed curved line indicates the 90\% completeness limit and the straight dotted line the 50\% limit. All magnitudes are in the Vega system and corrected for Galactic extinction.}
    \label{fig:CMD}
\end{figure*}

HST Advanced Camera for Surveys (ACS) images of BC6\footnote{\dataset[DOI: 10.17909/h12e-gb42]{http://dx.doi.org/10.17909/h12e-gb42}} were taken in February 2023 as part of the program GO-17267 (PI: M.~Jones). The imaging used the Wide Field Channel of ACS with the F606W and F814W filters. Over the course of two orbits, four exposures were taken in each filter, resulting in total integration times of approximately 2000~s in F606W and F814W. A false color composite of these images is shown in Figure~\ref{fig:HI_map} (right), centered on BC6.

Point source photometry was carried out with \texttt{DOLPHOT} \citep{Dolphin2000,Dolphin2016}, which uses the Vega magnitude system. Thus, our point source photometry is in the Vega system. Quality cuts to eliminate unwanted sources (e.g. background galaxies and diffraction spikes) were made as described in Section~3.1 of \citet{Jones+2022b}. Artificial star tests were used to estimate the 90\% and 50\% completeness limits of the source catalog \citep[as in][]{Jones+2022}. The \texttt{Python} package \texttt{dustmaps} \citep{dustmaps} was used to correct the point source photometry for Galactic extinction using the maps of \citet{Schlegel+1998} and the coefficients from \citet{Schlafly+2011} for the F606W and F814W filters. The typical extinction was very minor with a mean E(B-V) value of 0.024 mag. Stars within the ellipses in Figure~\ref{fig:HI_map} (right) are shown in the CMD in Figure~\ref{fig:CMD}. A background CMD is also shown for an equal area region far from BC6.

In addition to the point source photometry we also performed aperture photometry on the HST images in order to measure the integrated magnitude of BC6 in the F606W and F814W filters.
To do this we used the drizzled images in each filter and the apertures plotted in Figure~\ref{fig:HI_map} (right). Owing to the highly irregular morphology of BC6 these apertures were manually constructed to encompass the regions containing blue stars and UV emission, as well as with the distribution of light in the NGVS image (Figure~\ref{fig:BC6_NGVS}, left). Using the \texttt{Photutils} \citep{photutils} package the integrated magnitude within these apertures was measured and the background contribution subtracted based on the median sky value within annular apertures around each component. For comparison purposes we also convert to absolute V-band magnitude based on \citet{Sirianni+2005} and an assumed distance of 16.5~Mpc. These magnitude measurements are quoted in Table~\ref{tab:props}.

\subsection{Hobby-Eberly Telescope optical spectroscopy}
\label{sec:HET_data}

\begin{figure*}
    \centering
    \includegraphics[width=1.5\columnwidth]{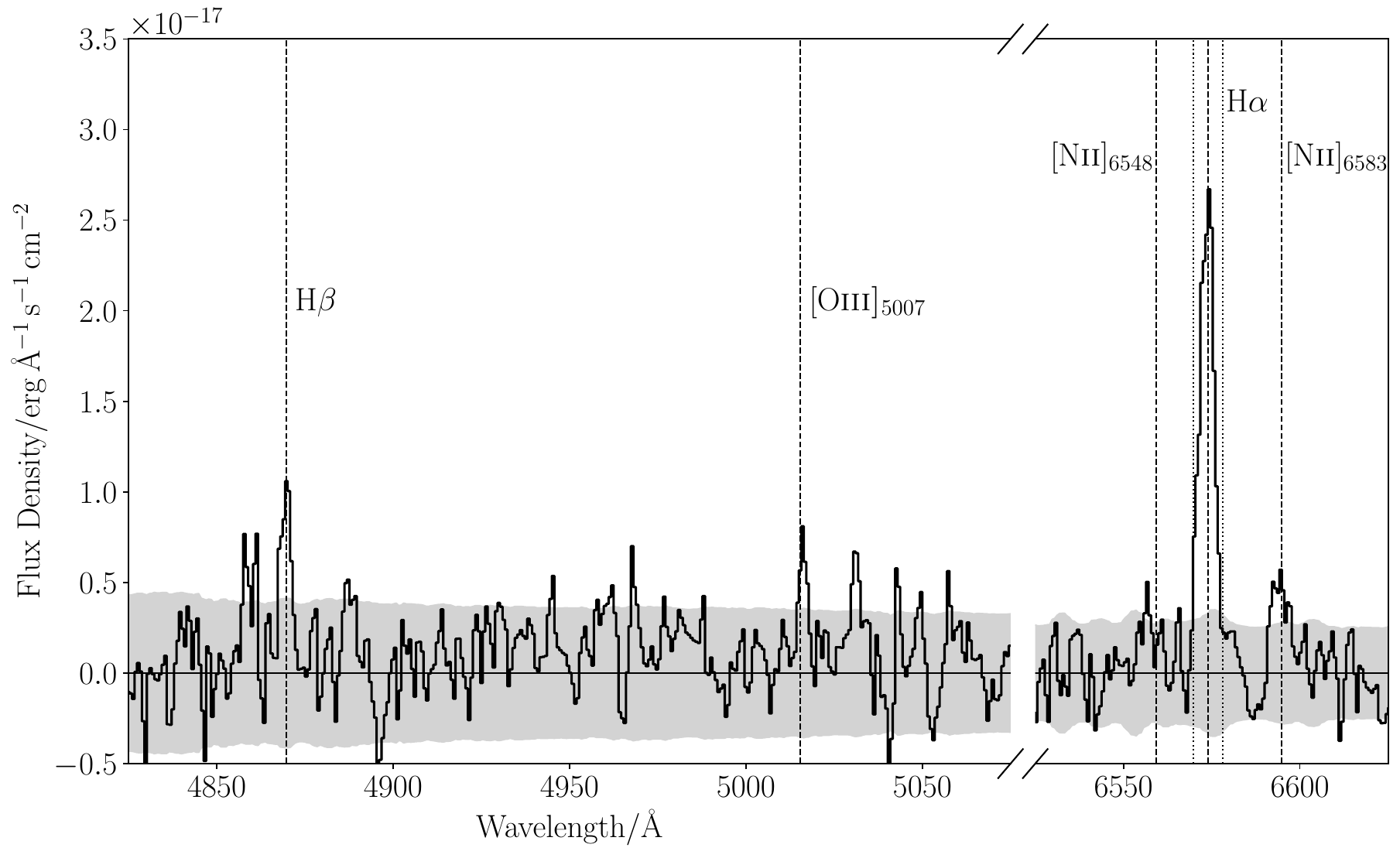}
    \caption{HET LRS2-B spectrum of BC6 showing the H$\alpha$, H$\beta$, [N{\sc ii}], and [O{\sc iii}] lines. The marked wavelengths of each line (vertical dashed lines) are all shifted to match the redshift of H$\alpha$. The vertical dotted lines indicate $\pm$2 standard deviations of the Gaussian fit to the H$\alpha$ line. This is also the window that was used to extract the line fluxes of all other lines (Table~\ref{tab:lines}).}
    \label{fig:HET_spec}
\end{figure*}

We observed the main component of BC6 (the largest ellipse in Figure~\ref{fig:HI_map}, right) with the Blue spectrograph of the Low Resolution Spectrograph 2 \citep[LRS2-B,][]{LRS2} on the 10m HET \citep{HET1,HET2} under clear dark skies on the night of April 19th, 2023 with a 3300~s exposure. The LRS2-B spectrograph has $R\approx1100$ and an integral field unit which is 6\arcsec$\times$12\arcsec \ with 0.6\arcsec \ lenslet fibers on sky, giving an abundance of sky fibers around the relatively compact optical emission from BC6. We use the H$\alpha$ line to determine which spaxels include emission from BC6 and extract all spaxels within 1.5\arcsec \ of the peak of the source with the \texttt{LRS2Multi} extension to the \texttt{Panacea} data reduction pipeline.\footnote{\url{https://github.com/grzeimann/LRS2Multi} and \url{https://github.com/grzeimann/Panacea}} Briefly, this data reduction includes the necessary detector characterizations, fiber extraction, wavelength calibration, mirror illumination correction, sky subtraction, and 1-D spectral source extraction including a treatment of differential atmospheric refraction in the sky and object spectra. The resulting spectrum is shown in Figure~\ref{fig:HET_spec}.

The spectrum is fairly typical of an \hii \ region, with clear, albeit relatively weak, emission lines. Only the H$\alpha$ line is high enough S/N to be fit with a Gaussian. Thus, this line alone was used to determine the redshift $cz = 500 \pm 4$~\kms \ (with a barycentric correction applied). To obtain fluxes from the other lines we extract the integrated signal within a $\pm$2$\sigma_{\mathrm{H}\alpha}$ region centered on the redshifted wavelength on each line. Here $\sigma_{\mathrm{H}\alpha}$ is the standard deviation of the Gaussian profile fit to the H$\alpha$ line. Flux errors are estimated by summing the uncertainty values in the same wavelength range and dividing by the square root of the number of channels included. 

A reddening correction was calculated by assuming an intrinsic value of $\mathrm{H}\alpha/\mathrm{H}\beta = 2.87$ \citep[for an \hii \ region at $10^4$~K;][]{Osterbrock}, an $R_V$ value of 3.1, and the \citet{Fitzpatrick2004} extinction law. Based on the observed line ratio $\mathrm{H}\alpha/\mathrm{H}\beta = 3.28$ we calculated an extinction value of $E(B-V) = 0.11$. This suggests that there is some internal extinction within BC6 as the \citet{Schlegel+1998} dust map in this direction indicates minimal Galactic extinction ($E(B-V) \approx 0.02$). Observed line fluxes were corrected for extinction \citep[using the extinction model of][]{Fitzpatrick2004} implemented in the \texttt{Python} package \texttt{dust\_extinction}.\footnote{\url{https://dust-extinction.readthedocs.io}}
These fluxes and uncertainties are provided in Table~\ref{tab:lines}. For consistency the quoted H$\alpha$ flux does not correspond to that calculated with the Gaussian fit, but was instead calculated in the same manner as the other lines.

We note here that the $E(B-V) = 0.11$ extinction estimate is strictly speaking only applicable to the \hii \ region as it is likely the result of dust in its immediate vicinity. For simplicity we apply the Galactic extinction correction only for all stars in the HST CMD (\S\ref{sec:HST_data} \& Figure~\ref{fig:CMD}), even though for a few stars in the immediate vicinity of the \hii \ region the higher estimate is likely more appropriate.

With the line flux measurements we used two line ratio metrics to estimate the gas-phase oxygen abundance in BC6 \citep[following][]{Pettini+2004}, N2=[N{\sc ii}]/H$\alpha$ and O3N2=([O{\sc iii}]/H$\beta$)/([N{\sc ii}]/H$\alpha$). We note that these ratios are designed to be almost independent of our extinction correction as [O{\sc iii}]$_{5007}$ and H$\beta$ occur at almost the same wavelength, as do [N{\sc ii}]$_{6583}$ and H$\alpha$. We see (in Table~\ref{tab:lines}) that both metrics give almost exactly the same oxygen abundance, despite each method having an intrinsic scatter of $\sim$0.2~dex.

Finally, we made a second estimate of the SFR based on the H$\alpha$ line flux. Following the standard \citet{Kennicutt1998} conversion this equates to a SFR of $\log (\mathrm{SFR}_{\mathrm{H}\alpha}/\mathrm{M_\odot \, yr^{-1}}) = -4.36 \pm 0.13$.

\section{Results} \label{sec:results}

\begin{table}[]
    \centering
    \caption{Properties of BC6}
    \hspace{-0.3in}
    \begin{tabular}{lc}
    \hline\hline
    Parameter       &  Value\\ \hline
    RA (J2000)              & 12:29:35 \\
    Dec. (J2000)            & $+$9:44:32 \\
    $v_{\mathrm{H}\alpha}$/\kms & $500\pm4$ \\
    $m_g$/mag (AB)          & $21.01\pm0.15$ \\
    $m_i$/mag (AB)          & $21.16\pm0.40$ \\
    $m_\mathrm{F606W}$/mag (AB)          & $20.69\pm0.15$ \\
    $m_\mathrm{F814W}$/mag (AB)          & $20.94\pm0.37$ \\
    $M_V$/mag (AB)          & $-10.4$\\
    $\log L_V/\mathrm{L_\odot}$ & 6.05\\
    $\log (\mathrm{SFR_{NUV}/M_\odot \, yr^{-1}})$ & $-3.51\pm0.04$ \\
    $\log (\mathrm{SFR}_{\mathrm{H}\alpha}/\mathrm{M_\odot \, yr^{-1}})$ & $-4.36 \pm 0.13$ \\
    $12 + \log(\mathrm{O/H})_\mathrm{O3N2}$ & $8.58 \pm 0.25$ \\
    $\log M_\ast$/\Msol  & $\sim$4.4 \\
    \hline
    \end{tabular}
    \tablenotetext{}{Absolute quantities assume a distance of 16.5~Mpc.}
    \label{tab:props}
\end{table}

\begin{table*}
\centering
\caption{Emission line fluxes and metallicity estimates of BC6}
\hspace{-0.7in}
\begin{tabular}{cccccc}
\hline\hline
H$\alpha$ & H$\beta$ &  [N{\sc ii}]$_{6583}$ & [O{\sc iii}]$_{5007}$ & $12 + \log(\mathrm{O/H})_\mathrm{N2}$ & $12 + \log(\mathrm{O/H})_\mathrm{O3N2}$         \\ \hline
$168 \pm 10$ & $59 \pm 14$ & $38 \pm 9$ & $38 \pm 12$ & $8.53 \pm 0.21$ & $8.58 \pm 0.25$ \\ \hline
\end{tabular}
\tablenotetext{}{Line fluxes are in units of $10^{-18} \; \mathrm{erg\,s^{-1}\,cm^{-2}}$ and are corrected for extinction.}
\label{tab:lines}
\end{table*}

\subsection{Morphology and stellar population} \label{sec:stellar_pop}

The optical appearance of BC6 is remarkably similar to that of BC4, which \citet{Jones+2022b} concluded was likely in the process of becoming gravitationally unbound after losing its \hi \ gas reservoir (presumably the majority of its initial total mass). Like BC4, BC6 is broken up into a few main components that are separated by 10s of arcsec (10\arcsec~$=800$~pc at 16.5~Mpc). Each component contains multiple clumps of blue stars and has associated UV emission. In the HST image these clumps of blue stars are partially resolved, with the brightest stars individually detected.

In the CMD of these resolved stars (Figure~\ref{fig:CMD}) we see a very similar stellar population to those of other blue blobs \citep{Sand+2017,Jones+2022,Jones+2022b}: a population of predominantly very blue (F606W-F814W = 0) stars, and a second population at approximately F606W-F814W = 1. As in our previous works, we argue that this population is consistent with very young main sequence stars and the blue and red sides of the Helium burning branch at the distance of the Virgo cluster (16.5~Mpc). The PARSEC \citep[PAdova
and TRieste Stellar Evolution Code;][]{Bressan+2012} isochrones overplotted in the middle panel of Figure~\ref{fig:CMD} indicate that the entirety of the observed population is consistent with stars in the age range of 10-100~Myr. In particular, we point out that these isochrones were not fitted to the CMD in any way, yet the color range of the Helium burning stars in the isochrones matches closely with the observed color range. This range is mainly determined by the metallicity of the stars, which we independently measured (and input to the isochrones) from optical spectroscopy of the \hii \ regions (\S\ref{sec:HET_data} \& \S\ref{sec:metals}).

We also see no evidence of an old population in the form of a red giant branch (RGB). However, at the distance of Virgo the tip of the RGB should occur at approximately F814W = 27 \citep{Jang+2017}, which is roughly the location of our 50\% completeness limit. Even so, if a significant underlying old population were present, we would expect to see a build up of RGB stars around the completeness limit. The absence of this population implies that BC6's stellar population is purely a young population, however, without deeper imaging it is impossible to determine the exact age of the oldest stars.

The two SFR estimates (Table~\ref{tab:props}) from NUV and H$\alpha$ differ by almost an order of magnitude. Although the HET observation only covered a single component of BC6, it was the main body (Figure~\ref{fig:HI_map}, right), where most of the UV flux originates. As NUV is sensitive to SF over a longer timescale than H$\alpha$, this may be an indication that SF is currently ramping down in BC6. However, given the very low SFR in BC6 it is also possible that this is merely indicative of stochasticity in the number of high mass stars produced in any given $\sim$10~Myr period.

\subsection{Metallicity} \label{sec:metals}

\citet{Beccari+2017} and \citet{Bellazzini+2022} measured the metallicities of five blue blobs (SECCO1, BC1, and BC3-5) using the same line ratio metrics as used here (see \S\ref{sec:HET_data}) and found a range of $8.29 < 12 + \log(\mathrm{O/H}) < 8.73$. BC6 fits comfortably within this range ($8.58 \pm 0.25$). Thus, as with its stellar population, it seems quite consistent with the established sample of blue blobs. 

Following the same rationale as \citet{Jones+2022b}, the above metallicity estimate implies that the progenitor galaxy of BC6 was likely in the stellar mass range $8.2 \lesssim \log(M_\ast/\mathrm{M_\odot}) \lesssim 10.2$, assuming it roughly follows the mass--metallicity relation \citep{Andrews+2013}. The broad range is a reflection of the large uncertainty in the metallicity. Unfortunately, over this stellar mass range the typical gas fraction ($M_\mathrm{HI}/M_\ast$) of a field galaxy smoothly transitions over the range $10 \gtrsim M_\mathrm{HI}/M_\ast \gtrsim 0.1$ \citep[e.g.][]{Huang+2012}, thus the fact that BC6 is accompanied by $\sim$10$^9$~\Msol \ of \hi \ gas does little to help narrow down the progenitor's stellar mass. However, it does mean that the progenitor must have lost a significant fraction, probably the majority, of its \hi \ gas reservoir.

\subsection{Stellar mass}

To estimate the stellar mass of BC6 we follow the same approach as \citet{Jones+2022b}, that is, assume a constant SFR and continuously build up a synthetic stellar population (of matching metallicity) until the observed total F814W magnitude of BC6 is reached. Using this approach we estimate a stellar mass of $\sim 2 \times 10^4$~\Msol, which at the observed (NUV-based) SFR could have been built up over approximately 80~Myr. If we instead use $g$ and $i$-band aperture photometry from the NGVS images and the scaling relations of \citet{Zibetti+2009} and \citet{Taylor+2011}, we obtain estimates of $5 \times 10^4$ and $1 \times 10^5$~\Msol, respectively.

It is encouraging that all these estimates fall within a single order of magnitude, even if they still vary by a factor of five. Although our initial estimate is smaller than both the others, we expect that this is the most representative of the actual stellar mass of BC6. Fundamentally, all three estimates rely on a mass-to-light scaling to produce a stellar mass, but only in our initial estimate is the synthetic population used to generate that mass-to-light ratio based on the observed metallicity and current SFR of BC6. This population is also quite atypical, with no apparent old population (older than a few hundred Myr), thus standard scaling relations are unlikely to be entirely appropriate.

With a stellar mass of $2 \times 10^4$~\Msol, BC6 must have a gas fraction ($M_\mathrm{HI}/M_\ast$) of around 3,000 if we only consider the gas in the coincident cloud (C3; Figure~\ref{fig:HI_map}, left), or over 20,000 if we include the whole \hi \ cloud complex. This makes it one of, if not the most \hi-rich extragalactic stellar systems known.

\section{Discussion} \label{sec:discuss}

\subsection{Point of origin} \label{sec:origin}

\begin{figure*}
    \centering
    \includegraphics[width=0.8\textwidth]{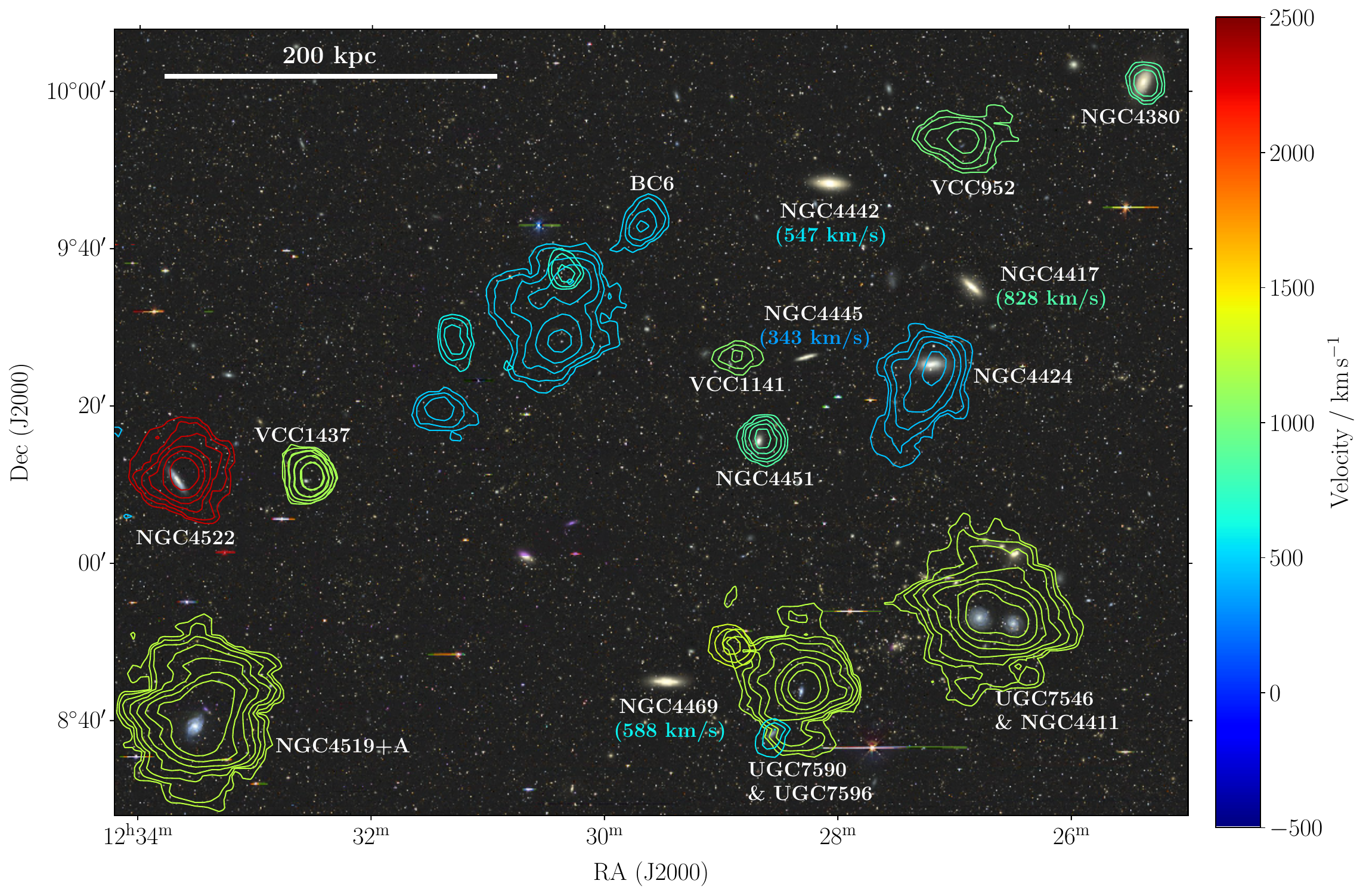}
    \caption{A wide-field DECaLS $grz$ image showing the galaxies in the vicinity of BC6 and the \hi \ complex. Contours showing \hi \ emission from ALFALFA are overlaid (same contour levels as Figure~\ref{fig:HI_map}). Each source identified by \texttt{SoFiA} is plotted separately and the color of the contours corresponds to the flux weighted mean velocity of each the source. Velocities for undetected sources are included as numerical values \citep[from][]{Cappellari+2011,SDSSDR13}. We note that although NGC~4445 is weakly detected by ALFALFA \citep{Haynes+2018}, it was not recovered in our automated extraction with \texttt{SoFiA} as it has lost most of its \hi.}
    \label{fig:widefield}
\end{figure*}

\begin{table*}
\centering
\caption{Candidate progenitor galaxies for BC6}
\hspace{-0.8in}
\label{tab:progen}
\begin{tabular}{ccccccccc}
\hline \hline
Galaxy & $v$/\kms & $\Delta v$/\kms & Separation & Direction & $\log M_\mathrm{HI}$ & $\log M_\mathrm{HI,pred}$  & $\log M_\mathrm{HI,lost}$& $12+\log(\mathrm{O/H})$ \\ \hline
NGC~4380 & 963  & 463  & 79\arcmin  & NW & 8.51 & 9.40 & 9.34 &     \\
NGC~4390 & 1101 & 601  & 86\arcmin  & NW & 8.68 & 8.85 & 8.36 & 8.37 \\
NGC~4424 & 437  & -63  & 47\arcmin  & W  & 8.34 & 9.43 & 9.39 & 8.50 \\
NGC~4442 & 547  & 47   & 37\arcmin  & NW & $<$7.8 & 9.61 & $>$9.60 &  \\
NGC~4445 & 353  & -147 & 31\arcmin  & W  & 7.71 & 9.19 & 9.17 &      \\
NGC~4519 & 1216 & 716  & 71\arcmin  & SE & 9.47 & 9.33 &  & 8.43 \\
NGC~4522 & 2329 & 1819 & 54\arcmin  & SE & 8.66 & 9.45 & 9.37 & 8.45 \\
NGC~4535 & 1963 & 1463 & 100\arcmin & SE & 9.57 & 9.93 & 9.68 & 8.52 \\ \hline
\end{tabular}
\tablenotetext{}{Columns: 1) Galaxy name. 2) Heliocentric velocity from ALFALFA \citep{Haynes+2018}. 3) Velocity difference between galaxy and BC6 (H$\alpha$ velocity). 4) Projected separation from the center of the \hi \ complex (taken as the approximate mid-point between C1 and C4). 5) Approximate direction from the center of the \hi \ complex to a given galaxy. 6) \hi \ mass in ALFALFA (in \Msol), typical uncertainties are 0.1~dex. NGC~4442 is undetected in ALFALFA and we therefore use the 50\% completeness limit \citep{Haynes+2011} and an assumed velocity width of 300~\kms \ to estimate an upper limit on its \hi \ mass. 7) Predicted original \hi \ mass in \Msol \ \citep{Jones+2018}, typical uncertainty is 0.21~dex. B-band diameters for the scaling relations were taken from \citet{RC3}. 8) The predicted \hi \ mass (in \Msol) lost (i.e. the linear difference between columns 6 and 7). 9) Metallicities measurements from \citet{DeVis+2019}, typical uncertainty is 0.14~dex.}
\end{table*}

Although the stellar counterpart, BC6, was unknown until now, the origin of the \hi \ cloud complex has been the subject of considerable discussion in the literature since it was first detected \citep{Kent+2007,Kent+2009,Sorgho+2017,Minchin+2019}. However, the gas-phase metallicity offers a new piece of pertinent information, and we also approach the question of the point of origin from a new standpoint in light of similar discussions for other blue blobs \citep{Adams+2015,Beccari+2017,Sand+2017,Bellazzini+2018,Calura+2020,Jones+2022,Jones+2022b}.

As mentioned by \citet{Minchin+2019}, tidal stripping does not seem to be a viable mechanism to form this structure. Not only is its isolation more difficult to explain with the low velocities ($\lesssim$300~\kms) involved in tidal stripping \citep[e.g.][]{Bournaud+2006}, but $10^9$~\Msol \ represents a significant amount of \hi \ gas for almost any galaxy and it is not possible to tidally remove a large fraction of a galaxy's gas disk without also stripping stars. The exception to this would be if the progenitor was a particularly massive and \hi-rich galaxy, however, that would also require a close passage from a similarly massive perturber to strip the gas, with neither remaining connected to the stripped gas in the present. Thus, the complete lack of stars over almost all of such a massive \hi \ complex is a strong argument for ram pressure stripping and we will only consider the formation of BC6 via this mechanism.

The \hi \ cloud complex (Figure~\ref{fig:HI_map}) is about 40\arcmin \ across ($\sim$190~kpc), making it plausible that the progenitor galaxy is a degree or more away. Unfortunately, despite there being no clear progenitor galaxy in the immediate vicinity of BC6, there are 10s of potential parent objects that are Virgo members within a few degrees (Figure~\ref{fig:widefield}). Both \citet{Kent+2009} and \citet{Minchin+2019} focused on NGC~4445 and NGC~4424 as these are only $\sim$30\arcmin \ to the SW of BC6 and have similar radial velocities to the gas complex \citep[346~\kms\ and 437~\kms, respectively;][]{SDSSDR13,Haynes+2018}. NGC~4424 is in the process of being ram pressure stripped \citep{Chung+2009,Sorgho+2017}, but its \hi \ tail points to the SE, roughly parallel to the \hi \ complex containing BC6, rather than towards it. This makes NGC~4424 an unlikely point of origin of the \hi \ complex. In the case of NGC~4445 the $M_\mathrm{HI}$--B-band diameter scaling relation of \citet{Jones+2018} indicates that its expected original \hi \ content (before falling into the Virgo cluster) would have been $\log M_\mathrm{HI}/\mathrm{M_\odot} = 9.19 \pm 0.21$ \citep[and it still contains $\log M_\mathrm{HI}/\mathrm{M_\odot} = 7.7$;][]{Haynes+2018}. The observed mass of the \hi \ complex therefore represents almost its entire original \hi \ content. In \citet{Sorgho+2017}, figure A1, NGC~4445 also appears to have a ram pressure tail pointing away from the \hi \ complex based on observations from the Karoo Array Telescope (KAT-7). However, \citet{Minchin+2021} did not detect this tail in deep Arecibo observations, suggesting that it was likely an artifact in the KAT-7 data. Thus, NGC~4445 remains a possible point of origin, with the caveat that the \hi \ mass of the complex represents virtually its entire original \hi \ content.

In simulations of high speed ram pressure stripping events \citep[e.g.][]{Kapferer+2009,Tonnesen+2012,Tonnesen+2021,Goller+2023}, on scales much larger than a galaxy, the tails that are formed are usually relatively linear features. This means that for progenitor candidates that are within $\sim$1$^\circ$, but not aligned with the major axis of the \hi \ complex, a larger than expected opening angle would be required to explain the geometry of the structure. It would perhaps still be possible to reproduce the observed geometry by involving clumpy structure within the ICM (the simulations assume the ICM is uniform and moving at a constant velocity). However, in an attempt to narrow our focus to the most likely candidates, we will now consider potential progenitors that are roughly  NW and SE of BC6, such that the cloud geometry can readily be explained by a simple ram pressure stripping scenario. There are six potentially viable progenitors within a few degrees: NGCs 4380, 4390, and 4442 to the NW, and NGCs 4519, 4522, and 4535 to the SE. All are massive enough to meet the mass--metallicity requirement from \S\ref{sec:metals}, but three (NGCs 4380, 4442, and 4535) are actually above the expected mass range \citep[stellar masses from][]{Mosenkov+2019}. All, except NGC~4442, currently contain some \hi, detected in ALFALFA. The basic properties of these galaxies are summarized in Table~\ref{tab:progen}.

We start by considering NGC~4442 as this is the closest major galaxy in projection (and velocity) along the direction of the major axis of the \hi \ complex. This galaxy is undetected in ALFALFA and its expected original \hi \ content indicates that it has likely lost enough gas to account for the \hi \ complex. However, this is also a relatively massive \citep[$\log M_\ast/\mathrm{M_\odot} = 10.6$;][]{Mosenkov+2019} and lenticular galaxy, and has probably been without \hi \ gas for some time. It seems quite improbable that a major stripping event could have occurred and NGC~4442 could have changed from a gas-rich galaxy to a quenched lenticular with no signs of the interaction, in the time it took for the gas to move a few hundred kpc away. We therefore do not consider NGC~4442 a likely point of origin.

We see from the metallicity measurements (Table~\ref{tab:progen}) that most of the galaxies with measurements could plausibly be matches for BC6 (Table~\ref{tab:lines}), with only NGC~4390 being significantly disfavored because of its metallicity. The predicted original \hi \ mass of NGC~4390 also suggests that it never would have had enough \hi \ to form the cloud complex associated with BC6. Furthermore, it appears NGC~4390 has not lost a significant amount of \hi.\footnote{We note that the predicted original \hi \ mass is usually intended as a statistical quantity and has considerable uncertainty for any individual galaxy. For example, the measured \hi \ content of NGC~4519 \textit{exceeds} its predicted original \hi \ mass by over 0.1~dex. These values should be considered a rough guide only.}

NGC~4380 was mapped in the VLA Imaging of Virgo in Atomic gas project \citep[VIVA;][]{Chung+2009}, which found that it has a truncated \hi \ disk. This indicates a past episode of ram pressure stripping, however, the direction of stripping is unknown. This galaxy is also above the stellar mass range expected for the progenitor \citep[$\log M_\ast/\mathrm{M_\odot} = 10.4$;][]{Mosenkov+2019}.

Unfortunately, VIVA did not map either NGC~4519 or NGC~4535. Both are detected in ALFALFA, but these data have an angular resolution of $\sim$4\arcmin. At this resolution NGC~4535 appears undisturbed, but the \hi \ distribution of NGC~4519 is clearly extended to the NW (towards BC6). However, NGC~4519 has a neighbor immediately to the NW (NGC~4519A) separated by only $\sim$200~\kms \ \citep{Binggeli+1985}. Thus the disturbance in the \hi \ distribution of NGC~4519 may be the result of either tides or ram pressure and with the resolution of the ALFALFA data there is little chance of distinguishing the two. 

NGC~4522 was mapped in VIVA and is a well-known example of a galaxy undergoing ram pressure stripping \citep[e.g.][]{Kenney+1999,Kenney+2004,Vollmer+2004,Chung+2009}. Furthermore, the direction of the stripping appears to be aligned along the direction towards BC6. However, BC6 and NGC~4522 are separated by $\sim$1800~km/s along the line-of-sight. \citet{Vollmer+2006} attempted to simulate the gas stripping occurring in NGC~4522, finding that it was best fit by a very high velocity relative to the ICM ($\sim$3500~\kms), with a velocity component in the SE direction of $\sim$3000~\kms. At this velocity the distance between NGC~4522 and BC6 could be traversed in $\sim$100~Myr. \citet{Vollmer+2006} estimate that the period of maximum stripping pressure on NGC~4522 was approximately 50~Myr ago. These two time frames do not exactly line up, but \citet{Vollmer+2006} indicated that significant stripping is expected to begin well before the peak ram pressure is reached. With these high velocities it may also be possible to explain the large velocity offset between BC6 and NGC~4522, if there is a significant component of NGC~4522's velocity (relative to the ICM) along the line-of-sight. Indeed, \citet{Minchin+2019} identified a low column density extension of NGC~4522's ram pressure tail pointing in the direction of BC6 and the \hi \ complex. Although, this makes it tempting to conclude that the two are likely connected, this only extends about 5\arcmin \ and 200~\kms \ towards the complex and therefore cannot be considered conclusive. Having said this, extrapolating this same velocity gradient another $\sim$40\arcmin \ would bring us to both the approximate position and velocity of the \hi \ complex. 

Another point in favor of NGC~4522 is that ram pressure stripping simulations \citep[e.g.][]{Tonnesen+2012,Lee+2022} show that most SF in stripped gas occurs either in the immediate vicinity of the galaxy or in the distant extremity of the tail. As BC6 is the only apparent stellar counterpart of the entire \hi \ complex (see Appendix~\ref{sec:appendix}) and is at the NW end of the tail, to be consistent with these simulations we would expect the point of origin to be to the SE, which NGC~4522 is.

\citet{Minchin+2019} consider NGC~4522 as a possible progenitor, but reject it on a couple of grounds. Firstly, they argue that if the \hi \ complex originated from NGC~4522 then you might expect cloud C5 to be at higher radial velocity than C3 (cf. Figure~\ref{fig:HI_map}), because it is closer to NGC~4522, but the reverse is true. However, the velocity dispersion of the whole structure is quite low and such an inconsistency might be explained by the two clouds being stripped from different sides of the galaxy (thus having different initial velocities). Also, the initial velocity relative to the ICM would be rapidly reduced, thus the original objection may not be valid if the \hi \ complex has been separated from NGC~4522 for long enough for the entire structure to come to rest relative to the ICM. 

The second issue raised by \citet{Minchin+2019} is that it is unlikely that the local ICM radial velocity in this region of Virgo is as low as 500~\kms \ (recall that in \S\ref{sec:search} we mentioned that the typical velocity of galaxies in this regions of Virgo is $\sim$1000~\kms). If this is not the case then it is not plausible that the \hi \ gas could have initially had a radial velocity of $\sim$2300~\kms \ and then slowed to $\sim$500~\kms. This is a more serious objection to NGC~4522 as the progenitor and is difficult to dismiss. 
The ram pressure stripping simulations of \citet{Vollmer+2006} are best able to reproduce the extraplanar gas distribution of NGC~4522 with an incidence angle of about 60$^\circ$ between the galaxy's motion and the ICM's. However, in this configuration the radial velocity of the ICM would be approximately 1000~\kms. An ICM velocity as low as 500~\kms \ could be possible in different configurations, but \citet{Vollmer+2006} found that such configurations would not reproduce the observed gas distribution.
We note that a similar (though less severe) objection regarding the ICM velocity applies to all the candidate progenitor galaxies to the NW and SE listed in Table~\ref{tab:progen}, as these all have radial velocities above 500~\kms. If one of these is indeed the origin of BC6, then this may be an indication that we do not understand the motions (and perhaps, sub-structure) of the ICM in this region of Virgo, and it might be necessary to re-visit the hydrodynamical modeling of the gas stripping in NGC~4522 to determine if other factors such as irregular motions and clumpiness in the ICM could reconcile these differences.

Finally, \citet{Minchin+2019} noted that the apparent gap in \hi \ between the ram pressure tail still connected to NGC~4522 and the \hi \ complex around BC6 would require two stripping episodes, with a break in between. However, simulations such as \citet{Kapferer+2009} and \citet{Tonnesen+2012} show similar morphology, with significant gaps in the \hi \ column density of tails despite continuous stripping. Gas that is initially heated (and therefore not visible in 21~cm line emission) may cool and condense back into \hi \ clouds in the tail, especially if the gas is metal-rich \citep[e.g.][]{Lee+2022}. However, it is not clear if this is a viable mechanism for such a large quantity of gas as is associated with BC6.

To summarize, we have considered a number of potential progenitor galaxies within a few degrees of BC6, but there is no clear parent object, in fact, all the candidates appear to have at least one disqualifying factor. We consider NGC~4522 to be the most likely parent object because it has similar metallicity and is a known case of extreme ram pressure stripping with the gas tail pointing towards BC6 both in projection and velocity. However, it is difficult to account for the kinematic separation of NGC~4522 and BC6. NGC~4380 and NGC~4519 also remain candidates for the progenitor, but this may simply be because there are fewer observations of these galaxies and we could be ignorant of disqualifying information. NGC~4445 was suggested as the point of origin by \citet{Kent+2009} and \citet{Minchin+2019}, however, we slightly disfavor this galaxy as the progenitor as the observed \hi \ complex represents almost the entirety of its estimated original gas content and the ram pressure tail would need a particularly wide opening angle to explain the extent of the complex.

\subsection{Fate}

Here we briefly consider the evaporation timescale for the clouds seen in the \hi \ complex, as a means to determine the fate of BC6. Following the discussion of \citet{Borthakur+2010}, which itself is based on \citet{Cowie+1977} and \citet{Vollmer+2001}, we see that cloud evaporation can proceed in either the classical or saturation scenario depending on the value of $\sigma_0$, the ratio of the classical and saturated heat flux through the spherical shell of the cloud, where
\begin{equation}
    \label{eq:heatflux}
    \sigma_0 \approx \frac{(T_\mathrm{ICM}/1\times10^7 \, \mathrm{K})^2}{n_\mathrm{ICM}R_\mathrm{cloud}}
\end{equation}
and $T_\mathrm{ICM}$ is the temperature of the hot ICM, $n_\mathrm{ICM}$ is its number density (in cm$^{-3}$), and $R_\mathrm{cloud}$ is the cloud radius (in pc). To evaluate this and proceed we need to know the physical properties of both a cloud within the \hi \ complex and the surrounding ICM.

To obtain approximate physical properties of a single cloud (as far as possible) we take the values for C2 (Figure~\ref{fig:HI_map}, left) which was observed with the VLA in \citet{Kent+2009}. In this higher resolution imaging the cloud has an elongated structure with two main concentrations of mass. For simplicity we take a quarter of the \hi \ major axis (1.5\arcmin) measured as the radius of a single cloud ($\sim$2~kpc) and half of the total \hi \ mass of C2 as its mass ($2.6\times10^7$~M$_{\odot}$). Here we have also multiplied by 1.4 to account for Helium and other metals. Assuming a spherical geometry gives a cloud number density of $\sim$0.3~$\mathrm{cm^{-3}}$.

For the ICM properties we use the simple $\beta$-model ICM density profile used in \citet{Vollmer+2001}, evaluated at 2.7$^\circ$ (the separation between BC6 and M~87), to obtain a ICM density estimate of $10^{-4}$~cm$^{-3}$. For the temperature we see from the profile plotted by \citet{Bohringer+1994} that at 2.7$^\circ$ from M~87 the average temperature of the ICM is roughly 2~keV ($1.5\times10^{7}$~K).

Using the above values in Equation~\ref{eq:heatflux}, we see that this cloud is in the saturated regime (i.e. $\sigma_0 > 1$) where the evaporation speed is limited by the rate heat can be conducted through the shell of the cloud. The expression derived by \citet{Borthakur+2010} for the lifetime of an \hi \ cloud in this regime is
\begin{equation}
    \label{eq:t_ev}
    \tau \approx \frac{n_\mathrm{cloud} R_\mathrm{cloud}}{n_\mathrm{ICM} \sqrt{T_\mathrm{ICM}}}  \; \mathrm{Myr}
\end{equation}
where $n_\mathrm{cloud}$ is the average number density of the cloud (in cm$^{-3}$). The other quantity are the same as in Equation~\ref{eq:heatflux}. Evaluating this expression gives a cloud lifetime for C2 of over a Gyr. 

Although this is only an order of magnitude estimate it is clear from the stellar population (\S\ref{sec:stellar_pop}) that the \hi \ complex has been traversing the ICM for at least $\sim$100~Myr and is still an extremely rich \hi \ structure. Furthermore, \citet{Calura+2020} considered in detail the survival of a much smaller gas cloud $M_\mathrm{HI} \sim 10^7$~\Msol\ (designed to mimic SECCO~1), finding that when simulating it traversing the ICM at high speed it could still survive as a distinct structure for around a Gyr. Thus, given the enormous reservoir still present, it seems that BC6 is not in imminent danger of losing its \hi \ gas and is unlikely to become unbound and disperse into the cluster in the short term.

However, as the \hi \ complex will not have a significant dark matter content, it unquestionably cannot survive indefinitely in the Virgo cluster. 
On long timescales it will eventually evaporate and its aging stars will disperse and contribute to the intra-cluster light.
Indeed, a Gyr is on the order of the orbital timescale of the cluster, thus it is possible that the complex would undergo interactions with inhomogeneities in the ICM within that time or, given its scale, might experience a strong tidal interaction with a galaxy. Either of which could act to disperse it and shorten its expected lifetime. In addition, if SF continues then stellar feedback may also act to disperse the structure, especially as it is not supported by the additional potential well of a dark matter halo. Thus, the evaporation timescale is likely an upper limit on how long this structure can survive, even if individual clouds might in theory survive that long.

\section{Conclusions} \label{sec:conclusions}

We have identified the stellar counterpart (BC6) of an enormous, isolated \hi \ structure ($\sim$190~kpc long, $M_\mathrm{HI} \sim 10^9$~\Msol) in the Virgo cluster that has been thought to be ``dark'' since its discovery by \citet{Kent+2007}. The extremely low (stellar) mass counterpart lies at the NW tip of the structure (Figure~\ref{fig:HI_map}). In ground-based imaging it appears as faint and irregular blue clumps, which HST images reveal to consist exclusively of young ($\lesssim$100~Myr) stars. Given this population, the total stellar mass of BC6 is likely only a few times $10^4$~\Msol, making it a truly extreme object in terms of gas fraction.

An optical spectrum of this counterpart from HET returned an H$\alpha$ redshift (500~\kms) consistent with the velocity of the co-spatial \hi \ gas seen in ALFALFA and with GBT observations, thereby confirming association. This spectrum also revealed that BC6 is metal-rich (given its low stellar mass), indicating that its gas complex must have originated in a galaxy where the gas could be pre-enriched. 

Given these properties, BC6 appears to be a member of the novel stellar systems described by \citet{Jones+2022b}, known as ``blue blobs," standing out only because of its extraordinary \hi \ content. The origin of this gas is as challenging to conclusively determine as are the origins of previously studied blue blobs. However, high speed ram pressure stripping remains the most plausible explanation, regardless of which cluster member was the origin. We favor NGC~4522 as a candidate point of origin as it has a similar metallicity, is a known example of ongoing extreme ram pressure stripping, and probably passed through the local peak of X-ray emission (from the ICM), where BC6 currently resides, about 100~Myr ago \citep{Bohringer+1994,Vollmer+2001}. However, this association is far from conclusive, in particular the kinematic separation of BC6 and NGC~4522 is difficult to explain.

Given how long this \hi \ structure was known without a stellar counterpart being discovered, it seems likely other ``dark'' \hi \ structures in Virgo \citep[e.g.][]{Kent+2007,Taylor+2012} may host similarly faint blue blob counterparts as well. This is a possibility that we aim to investigate through our ongoing citizen science search covering the entire cluster.

\begin{acknowledgments}
We thanks the anonymous referee for their constructive comments which helped to improve this work. We gratefully acknowledge the work of all the Zooniverse volunteers who have participated in the Blobs and Blurs project. This publication uses data generated via the Zooniverse.org platform, development of which is funded by generous support, including a Global Impact Award from Google, and by a grant from the Alfred P. Sloan Foundation. This work is based on observations made with the NASA/ESA Hubble Space Telescope, obtained at the Space Telescope Science Institute, which is operated by the Association of Universities for Research in Astronomy, Inc., under NASA contract NAS5-26555.  These observations are associated with program \# HST-GO-17267.  Support for program \# HST-GO-17267 was provided by NASA through a grant from the Space Telescope Science Institute.
DJS and the Arizona team acknowledges support from NSF grant AST-2205863.
Supported in part through the Arizona NASA Space Grant Consortium, Cooperative Agreement 80NSSC20M0041.
KS acknowledges support from the Natural Sciences and Engineering Research Council of Canada (NSERC).
DC acknowledges support from NSF grant AST-1814208.
AK acknowledges support from NSERC, the University of Toronto Arts \& Science Postdoctoral Fellowship program, and the Dunlap Institute.
We acknowledge the Texas Advanced Computing Center (TACC) at The University of Texas at Austin for providing high performance computing, visualization, and storage resources that have contributed to the results reported within this paper.
This work used images from the Dark Energy Camera Legacy Survey (DECaLS; Proposal ID 2014B-0404; PIs: David Schlegel and Arjun Dey). Full acknowledgment at \url{https://www.legacysurvey.org/acknowledgment/}.
This research used the facilities of the Canadian Astronomy Data Centre operated by the National Research Council of Canada with the support of the Canadian Space Agency. 
The Green Bank Observatory is a facility of the National Science Foundation operated under cooperative agreement by Associated Universities, Inc. The National Radio Astronomy Observatory is a facility of the National Science Foundation operated under cooperative agreement by Associated Universities, Inc.   
\end{acknowledgments}

%

\vspace{5mm}
\facilities{HST(ACS), GBT, HET, Arecibo, CFHT, CTIO, GALEX}


\software{\href{http://americano.dolphinsim.com/dolphot/}{\texttt{DOLPHOT}} \citep{Dolphin2000,Dolphin2016}, \href{https://gitlab.com/SoFiA-Admin/SoFiA}{\texttt{SoFiA}} \citep{SoFiA,Serra+2015}, \href{https://www.astropy.org/index.html}{\texttt{astropy}} \citep{astropy2013,astropy2018}, \href{https://reproject.readthedocs.io/en/stable/}{\texttt{reproject}} \citep{reproject}, \href{https://sites.google.com/cfa.harvard.edu/saoimageds9}{\texttt{DS9}} \citep{DS9}, \href{https://aladin.u-strasbg.fr/}{\texttt{Aladin}} \citep{Aladin2000,Aladin2014}, \href{https://casa.nrao.edu/}{\texttt{CASA}} \citep{CASA}, \href{https://gbtidl.nrao.edu/}{\texttt{GBTIDL}}, \href{https://dustmaps.readthedocs.io/en/latest/}{\texttt{dustmaps}} \citep{dustmaps}, \href{https://matplotlib.org/}{\texttt{matplotlib}} \citep{matplotlib}, \href{https://numpy.org/}{\texttt{numpy}} \citep{numpy}, \href{https://scipy.org/}{\texttt{scipy}} \citep{scipy1,scipy2}, \href{https://pandas.pydata.org/}{\texttt{pandas}} \citep{pandas1,pandas2}, \href{https://github.com/grzeimann/Panacea}{\texttt{Panacea}}, \href{https://github.com/grzeimann/LRS2Multi}{\texttt{LRS2Multi}}, \href{https://dust-extinction.readthedocs.io}{\texttt{dust\_extinction}}.}



\appendix

\section{Other counterpart candidates} \label{sec:appendix}

\begin{figure}
    \centering
    \includegraphics[width=0.5\columnwidth]{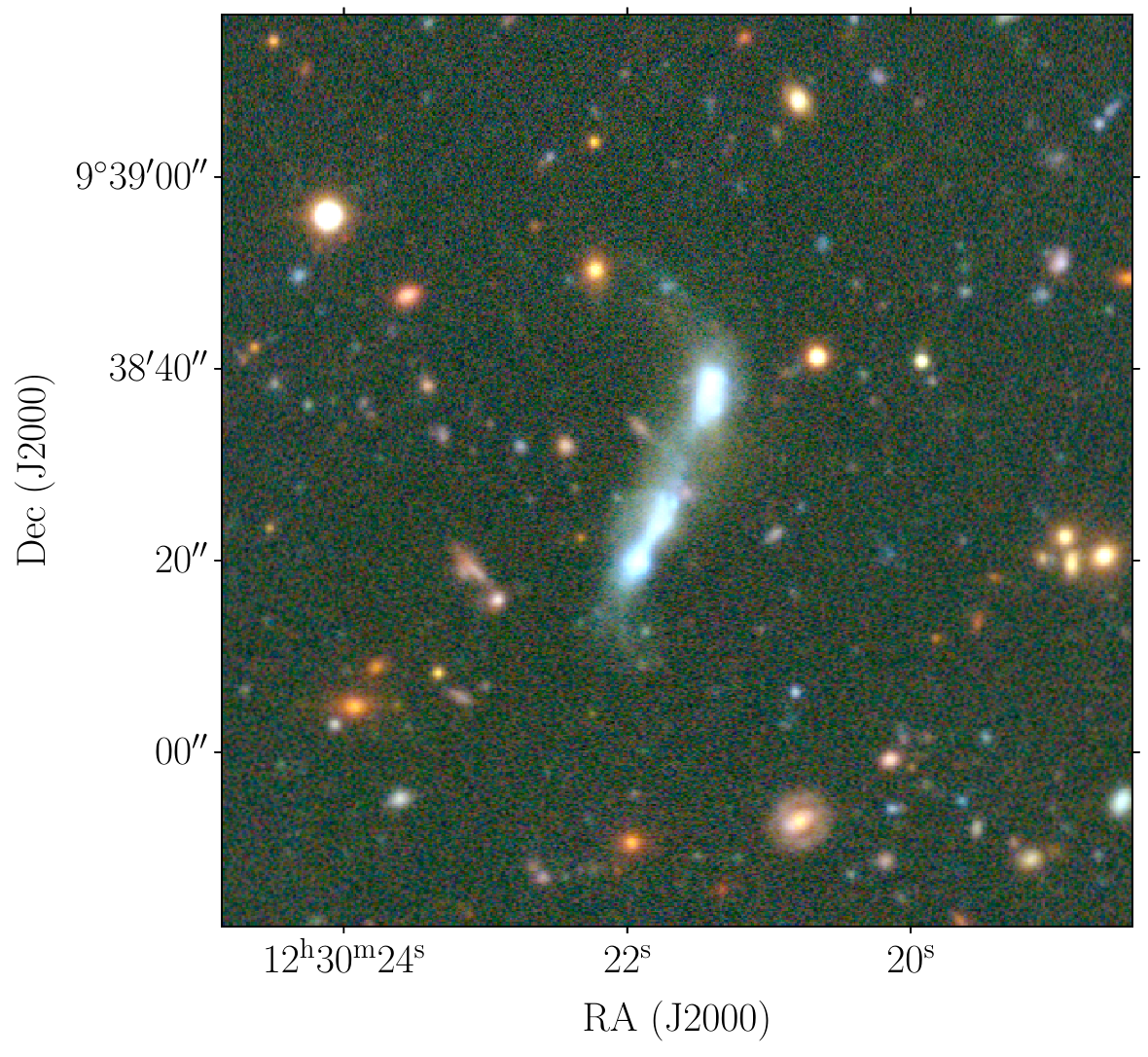}
    \caption{NGVS $ugi$ image of an additional blue blob candidate co-spatial with \hi \ cloud C4 (Figure~\ref{fig:HI_map}) identified in our citizen science search. However, this candidate appears to be background galaxies.}
    \label{fig:add_cand}
\end{figure}

Our citizen science search for additional blue blob candidates throughout Virgo was recently completed, we extracted all the candidates in this region to identify if BC6 is the only object associated with the \hi \ tail, or if there might be other candidate blue blobs coincident with other clouds.

We only found one additional candidate (Figure~\ref{fig:add_cand}) coincident with the \hi \ structure, next to cloud C4 (Figure~\ref{fig:HI_map}). The appearance of this candidate suggests that it is likely a background group of interacting galaxies. Although there are blue clumps, as in genuine blue blobs, they are surrounded by diffuse yellow light, probably made up of old stars that are very far away. Thus, we do not consider this likely to be a genuine blue blob or associated with the \hi \ structure, however, we plan to obtain optical spectroscopic follow-up in the future to rule this out.


\bibliography{refs}{}

\begin{thebibliography}{}
\expandafter\ifx\csname natexlab\endcsname\relax\def\natexlab#1{#1}\fi
\providecommand{\url}[1]{\href{#1}{#1}}
\providecommand{\dodoi}[1]{doi:~\href{http://doi.org/#1}{\nolinkurl{#1}}}
\providecommand{\doeprint}[1]{\href{http://ascl.net/#1}{\nolinkurl{http://ascl.net/#1}}}
\providecommand{\doarXiv}[1]{\href{https://arxiv.org/abs/#1}{\nolinkurl{https://arxiv.org/abs/#1}}}

\bibitem[{{Adams} {et~al.}(2013){Adams}, {Giovanelli}, \& {Haynes}}]{Adams+2013}
{Adams}, E. A.~K., {Giovanelli}, R., \& {Haynes}, M.~P. 2013, \apj, 768, 77, \dodoi{10.1088/0004-637X/768/1/77}

\bibitem[{{Adams} {et~al.}(2015){Adams}, {Cannon}, {Rhode}, {Janesh}, {Janowiecki}, {Leisman}, {Giovanelli}, {Haynes}, {Oosterloo}, {Salzer}, \& {Zaidi}}]{Adams+2015}
{Adams}, E.~A.~K., {Cannon}, J.~M., {Rhode}, K.~L., {et~al.} 2015, \aap, 580, A134, \dodoi{10.1051/0004-6361/201526857}

\bibitem[{{Albareti} {et~al.}(2017){Albareti}, {Allende Prieto}, {Almeida}, {Anders}, {Anderson}, {Andrews}, {Arag{\'o}n-Salamanca}, {Argudo-Fern{\'a}ndez}, {Armengaud}, {Aubourg}, {Avila-Reese}, {Badenes}, {Bailey}, {Barbuy}, {Barger}, {Barrera-Ballesteros}, {Bartosz}, {Basu}, {Bates}, {Battaglia}, {Baumgarten}, {Baur}, {Bautista}, {Beers}, {Belfiore}, {Bershady}, {Bertran de Lis}, {Bird}, {Bizyaev}, {Blanc}, {Blanton}, {Blomqvist}, {Bolton}, {Borissova}, {Bovy}, {Brandt}, {Brinkmann}, {Brownstein}, {Bundy}, {Burtin}, {Busca}, {Camacho Chavez}, {Cano D{\'\i}az}, {Cappellari}, {Carrera}, {Chen}, {Cherinka}, {Cheung}, {Chiappini}, {Chojnowski}, {Chuang}, {Chung}, {Cirolini}, {Clerc}, {Cohen}, {Comerford}, {Comparat}, {Correa do Nascimento}, {Cousinou}, {Covey}, {Crane}, {Croft}, {Cunha}, {Darling}, {Davidson}, {Dawson}, {Da Costa}, {Da Silva Ilha}, {Deconto Machado}, {Delubac}, {De Lee}, {De la Macorra}, {De la Torre}, {Diamond-Stanic}, {Donor}, {Downes}, {Drory}, {Du}, {Du Mas des Bourboux}, {Dwelly},
  {Ebelke}, {Eigenbrot}, {Eisenstein}, {Elsworth}, {Emsellem}, {Eracleous}, {Escoffier}, {Evans}, {Falc{\'o}n-Barroso}, {Fan}, {Favole}, {Fernandez-Alvar}, {Fernandez-Trincado}, {Feuillet}, {Fleming}, {Font-Ribera}, {Freischlad}, {Frinchaboy}, {Fu}, {Gao}, {Garcia}, {Garcia-Dias}, {Garcia-Hern{\'a}ndez}, {Garcia P{\'e}rez}, {Gaulme}, {Ge}, {Geisler}, {Gillespie}, {Gil Marin}, {Girardi}, {Goddard}, {Gomez Maqueo Chew}, {Gonzalez-Perez}, {Grabowski}, {Green}, {Grier}, {Grier}, {Guo}, {Guy}, {Hagen}, {Hall}, {Harding}, {Harley}, {Hasselquist}, {Hawley}, {Hayes}, {Hearty}, {Hekker}, {Hernandez Toledo}, {Ho}, {Hogg}, {Holley-Bockelmann}, {Holtzman}, {Holzer}, {Hu}, {Huber}, {Hutchinson}, {Hwang}, {Ibarra-Medel}, {Ivans}, {Ivory}, {Jaehnig}, {Jensen}, {Johnson}, {Jones}, {Jullo}, {Kallinger}, {Kinemuchi}, {Kirkby}, {Klaene}, {Kneib}, {Kollmeier}, {Lacerna}, {Lane}, {Lang}, {Laurent}, {Law}, {Leauthaud}, {Le Goff}, {Li}, {Li}, {Li}, {Li}, {Liang}, {Liang}, {Lima}, {Lin}, {Lin}, {Lin}, {Liu}, {Long}, {Lucatello},
  {MacDonald}, {MacLeod}, {Mackereth}, {Mahadevan}, {Maia}, {Maiolino}, {Majewski}, {Malanushenko}, {Malanushenko}, {Mallmann}, {Manchado}, {Maraston}, {Marques-Chaves}, {Martinez Valpuesta}, {Masters}, {Mathur}, {McGreer}, {Merloni}, {Merrifield}, {M{\'e}sz{\'a}ros}, {Meza}, {Miglio}, {Minchev}, {Molaverdikhani}, {Montero-Dorta}, {Mosser}, {Muna}, {Myers}, {Nair}, {Nandra}, {Ness}, {Newman}, {Nichol}, {Nidever}, {Nitschelm}, {O'Connell}, {Oravetz}, {Oravetz}, {Pace}, {Padilla}, {Palanque-Delabrouille}, {Pan}, {Parejko}, {Paris}, {Park}, {Peacock}, {Peirani}, {Pellejero-Ibanez}, {Penny}, {Percival}, {Percival}, {Perez-Fournon}, {Petitjean}, {Pieri}, {Pinsonneault}, {Pisani}, {Prada}, {Prakash}, {Price-Jones}, {Raddick}, {Rahman}, {Raichoor}, {Barboza Rembold}, {Reyna}, {Rich}, {Richstein}, {Ridl}, {Riffel}, {Riffel}, {Rix}, {Robin}, {Rockosi}, {Rodr{\'\i}guez-Torres}, {Rodrigues}, {Roe}, {Roman Lopes}, {Rom{\'a}n-Z{\'u}{\~n}iga}, {Ross}, {Rossi}, {Ruan}, {Ruggeri}, {Runnoe}, {Salazar-Albornoz}, {Salvato},
  {Sanchez}, {Sanchez}, {Sanchez-Gallego}, {Santiago}, {Schiavon}, {Schimoia}, {Schlafly}, {Schlegel}, {Schneider}, {Sch{\"o}nrich}, {Schultheis}, {Schwope}, {Seo}, {Serenelli}, {Sesar}, {Shao}, {Shetrone}, {Shull}, {Silva Aguirre}, {Skrutskie}, {Slosar}, {Smith}, {Smith}, {Sobeck}, {Somers}, {Souto}, {Stark}, {Stassun}, {Steinmetz}, {Stello}, {Storchi Bergmann}, {Strauss}, {Streblyanska}, {Stringfellow}, {Suarez}, {Sun}, {Taghizadeh-Popp}, {Tang}, {Tao}, {Tayar}, {Tembe}, {Thomas}, {Tinker}, {Tojeiro}, {Tremonti}, {Troup}, {Trump}, {Unda-Sanzana}, {Valenzuela}, {Van den Bosch}, {Vargas-Maga{\~n}a}, {Vazquez}, {Villanova}, {Vivek}, {Vogt}, {Wake}, {Walterbos}, {Wang}, {Wang}, {Weaver}, {Weijmans}, {Weinberg}, {Westfall}, {Whelan}, {Wilcots}, {Wild}, {Williams}, {Wilson}, {Wood-Vasey}, {Wylezalek}, {Xiao}, {Yan}, {Yang}, {Ybarra}, {Yeche}, {Yuan}, {Zakamska}, {Zamora}, {Zasowski}, {Zhang}, {Zhao}, {Zhao}, {Zheng}, {Zheng}, {Zhou}, {Zhu}, {Zinn}, \& {Zou}}]{SDSSDR13}
{Albareti}, F.~D., {Allende Prieto}, C., {Almeida}, A., {et~al.} 2017, \apjs, 233, 25, \dodoi{10.3847/1538-4365/aa8992}

\bibitem[{{Andrews} \& {Martini}(2013)}]{Andrews+2013}
{Andrews}, B.~H., \& {Martini}, P. 2013, \apj, 765, 140, \dodoi{10.1088/0004-637X/765/2/140}

\bibitem[{{Astropy Collaboration} {et~al.}(2013){Astropy Collaboration}, {Robitaille}, {Tollerud}, {Greenfield}, {Droettboom}, {Bray}, {Aldcroft}, {Davis}, {Ginsburg}, {Price-Whelan}, {Kerzendorf}, {Conley}, {Crighton}, {Barbary}, {Muna}, {Ferguson}, {Grollier}, {Parikh}, {Nair}, {Unther}, {Deil}, {Woillez}, {Conseil}, {Kramer}, {Turner}, {Singer}, {Fox}, {Weaver}, {Zabalza}, {Edwards}, {Azalee Bostroem}, {Burke}, {Casey}, {Crawford}, {Dencheva}, {Ely}, {Jenness}, {Labrie}, {Lim}, {Pierfederici}, {Pontzen}, {Ptak}, {Refsdal}, {Servillat}, \& {Streicher}}]{astropy2013}
{Astropy Collaboration}, {Robitaille}, T.~P., {Tollerud}, E.~J., {et~al.} 2013, \aap, 558, A33, \dodoi{10.1051/0004-6361/201322068}

\bibitem[{{Astropy Collaboration} {et~al.}(2018){Astropy Collaboration}, {Price-Whelan}, {Sip{\H{o}}cz}, {G{\"u}nther}, {Lim}, {Crawford}, {Conseil}, {Shupe}, {Craig}, {Dencheva}, {Ginsburg}, {VanderPlas}, {Bradley}, {P{\'e}rez-Su{\'a}rez}, {de Val-Borro}, {Aldcroft}, {Cruz}, {Robitaille}, {Tollerud}, {Ardelean}, {Babej}, {Bach}, {Bachetti}, {Bakanov}, {Bamford}, {Barentsen}, {Barmby}, {Baumbach}, {Berry}, {Biscani}, {Boquien}, {Bostroem}, {Bouma}, {Brammer}, {Bray}, {Breytenbach}, {Buddelmeijer}, {Burke}, {Calderone}, {Cano Rodr{\'\i}guez}, {Cara}, {Cardoso}, {Cheedella}, {Copin}, {Corrales}, {Crichton}, {D'Avella}, {Deil}, {Depagne}, {Dietrich}, {Donath}, {Droettboom}, {Earl}, {Erben}, {Fabbro}, {Ferreira}, {Finethy}, {Fox}, {Garrison}, {Gibbons}, {Goldstein}, {Gommers}, {Greco}, {Greenfield}, {Groener}, {Grollier}, {Hagen}, {Hirst}, {Homeier}, {Horton}, {Hosseinzadeh}, {Hu}, {Hunkeler}, {Ivezi{\'c}}, {Jain}, {Jenness}, {Kanarek}, {Kendrew}, {Kern}, {Kerzendorf}, {Khvalko}, {King}, {Kirkby}, {Kulkarni},
  {Kumar}, {Lee}, {Lenz}, {Littlefair}, {Ma}, {Macleod}, {Mastropietro}, {McCully}, {Montagnac}, {Morris}, {Mueller}, {Mumford}, {Muna}, {Murphy}, {Nelson}, {Nguyen}, {Ninan}, {N{\"o}the}, {Ogaz}, {Oh}, {Parejko}, {Parley}, {Pascual}, {Patil}, {Patil}, {Plunkett}, {Prochaska}, {Rastogi}, {Reddy Janga}, {Sabater}, {Sakurikar}, {Seifert}, {Sherbert}, {Sherwood-Taylor}, {Shih}, {Sick}, {Silbiger}, {Singanamalla}, {Singer}, {Sladen}, {Sooley}, {Sornarajah}, {Streicher}, {Teuben}, {Thomas}, {Tremblay}, {Turner}, {Terr{\'o}n}, {van Kerkwijk}, {de la Vega}, {Watkins}, {Weaver}, {Whitmore}, {Woillez}, {Zabalza}, \& {Astropy Contributors}}]{astropy2018}
{Astropy Collaboration}, {Price-Whelan}, A.~M., {Sip{\H{o}}cz}, B.~M., {et~al.} 2018, \aj, 156, 123, \dodoi{10.3847/1538-3881/aabc4f}

\bibitem[{{Beccari} {et~al.}(2017){Beccari}, {Bellazzini}, {Magrini}, {Coccato}, {Cresci}, {Fraternali}, {de Zeeuw}, {Husemann}, {Ibata}, {Battaglia}, {Martin}, {Testa}, {Perina}, \& {Correnti}}]{Beccari+2017}
{Beccari}, G., {Bellazzini}, M., {Magrini}, L., {et~al.} 2017, \mnras, 465, 2189, \dodoi{10.1093/mnras/stw2874}

\bibitem[{{Bellazzini} {et~al.}(2015){Bellazzini}, {Magrini}, {Mucciarelli}, {Beccari}, {Ibata}, {Battaglia}, {Martin}, {Testa}, {Fumana}, {Marchetti}, {Correnti}, \& {Fraternali}}]{Bellazzini+2015}
{Bellazzini}, M., {Magrini}, L., {Mucciarelli}, A., {et~al.} 2015, \apjl, 800, L15, \dodoi{10.1088/2041-8205/800/1/L15}

\bibitem[{{Bellazzini} {et~al.}(2018){Bellazzini}, {Armillotta}, {Perina}, {Magrini}, {Cresci}, {Beccari}, {Battaglia}, {Fraternali}, {de Zeeuw}, {Martin}, {Calura}, {Ibata}, {Coccato}, {Testa}, \& {Correnti}}]{Bellazzini+2018}
{Bellazzini}, M., {Armillotta}, L., {Perina}, S., {et~al.} 2018, \mnras, 476, 4565, \dodoi{10.1093/mnras/sty467}

\bibitem[{{Bellazzini} {et~al.}(2022){Bellazzini}, {Magrini}, {Jones}, {Sand}, {Beccari}, {Cresci}, {Spekkens}, {Karunakaran}, {Adams}, {Zaritsky}, {Battaglia}, {Seth}, {Cannon}, {Fuson}, {Inoue}, {Mutlu-Pakdil}, {Guhathakurta}, {Mu{\~n}oz}, {Bennet}, {Crnojevi{\'c}}, {Caldwell}, {Strader}, \& {Toloba}}]{Bellazzini+2022}
{Bellazzini}, M., {Magrini}, L., {Jones}, M.~G., {et~al.} 2022, \apj, 935, 50, \dodoi{10.3847/1538-4357/ac7c6d}

\bibitem[{{Binggeli} {et~al.}(1985){Binggeli}, {Sandage}, \& {Tammann}}]{Binggeli+1985}
{Binggeli}, B., {Sandage}, A., \& {Tammann}, G.~A. 1985, \aj, 90, 1681, \dodoi{10.1086/113874}

\bibitem[{{Binggeli} {et~al.}(1987){Binggeli}, {Tammann}, \& {Sandage}}]{Binggeli+1987}
{Binggeli}, B., {Tammann}, G.~A., \& {Sandage}, A. 1987, \aj, 94, 251, \dodoi{10.1086/114467}

\bibitem[{{Boch} \& {Fernique}(2014)}]{Aladin2014}
{Boch}, T., \& {Fernique}, P. 2014, in Astronomical Society of the Pacific Conference Series, Vol. 485, Astronomical Data Analysis Software and Systems XXIII, ed. N.~{Manset} \& P.~{Forshay}, 277

\bibitem[{{B{\"o}hringer} {et~al.}(1994){B{\"o}hringer}, {Briel}, {Schwarz}, {Voges}, {Hartner}, \& {Tr{\"u}mper}}]{Bohringer+1994}
{B{\"o}hringer}, H., {Briel}, U.~G., {Schwarz}, R.~A., {et~al.} 1994, \nat, 368, 828, \dodoi{10.1038/368828a0}

\bibitem[{{Bonnarel} {et~al.}(2000){Bonnarel}, {Fernique}, {Bienaym{\'e}}, {Egret}, {Genova}, {Louys}, {Ochsenbein}, {Wenger}, \& {Bartlett}}]{Aladin2000}
{Bonnarel}, F., {Fernique}, P., {Bienaym{\'e}}, O., {et~al.} 2000, \aaps, 143, 33, \dodoi{10.1051/aas:2000331}

\bibitem[{{Borthakur} {et~al.}(2010){Borthakur}, {Yun}, \& {Verdes-Montenegro}}]{Borthakur+2010}
{Borthakur}, S., {Yun}, M.~S., \& {Verdes-Montenegro}, L. 2010, \apj, 710, 385, \dodoi{10.1088/0004-637X/710/1/385}

\bibitem[{{Bournaud} \& {Duc}(2006)}]{Bournaud+2006}
{Bournaud}, F., \& {Duc}, P.~A. 2006, \aap, 456, 481, \dodoi{10.1051/0004-6361:20065248}

\bibitem[{Bradley {et~al.}(2020)Bradley, Sip{\H o}cz, Robitaille, Tollerud, Vin{\'{\i}}cius, Deil, Barbary, Wilson, Busko, G{\"u}nther, Cara, Conseil, Bostroem, Droettboom, Bray, Bratholm, Lim, Barentsen, Craig, Pascual, Perren, Greco, Donath, de~Val-Borro, Kerzendorf, Bach, Weaver, D'Eugenio, Souchereau, \& Ferreira}]{photutils}
Bradley, L., Sip{\H o}cz, B., Robitaille, T., {et~al.} 2020, astropy/photutils: 1.0.0, 1.0.0,  Zenodo, \dodoi{10.5281/zenodo.4044744}

\bibitem[{{Bressan} {et~al.}(2012){Bressan}, {Marigo}, {Girardi}, {Salasnich}, {Dal Cero}, {Rubele}, \& {Nanni}}]{Bressan+2012}
{Bressan}, A., {Marigo}, P., {Girardi}, L., {et~al.} 2012, \mnras, 427, 127, \dodoi{10.1111/j.1365-2966.2012.21948.x}

\bibitem[{{Burkhart} \& {Loeb}(2016)}]{Burkhart+2016}
{Burkhart}, B., \& {Loeb}, A. 2016, \apjl, 824, L7, \dodoi{10.3847/2041-8205/824/1/L7}

\bibitem[{{Calura} {et~al.}(2020){Calura}, {Bellazzini}, \& {D'Ercole}}]{Calura+2020}
{Calura}, F., {Bellazzini}, M., \& {D'Ercole}, A. 2020, \mnras, 499, 5873, \dodoi{10.1093/mnras/staa3133}

\bibitem[{{Cappellari} {et~al.}(2011){Cappellari}, {Emsellem}, {Krajnovi{\'c}}, {McDermid}, {Scott}, {Verdoes Kleijn}, {Young}, {Alatalo}, {Bacon}, {Blitz}, {Bois}, {Bournaud}, {Bureau}, {Davies}, {Davis}, {de Zeeuw}, {Duc}, {Khochfar}, {Kuntschner}, {Lablanche}, {Morganti}, {Naab}, {Oosterloo}, {Sarzi}, {Serra}, \& {Weijmans}}]{Cappellari+2011}
{Cappellari}, M., {Emsellem}, E., {Krajnovi{\'c}}, D., {et~al.} 2011, \mnras, 413, 813, \dodoi{10.1111/j.1365-2966.2010.18174.x}

\bibitem[{{Chengalur} {et~al.}(1995){Chengalur}, {Giovanelli}, \& {Haynes}}]{Chengalur+1995}
{Chengalur}, J.~N., {Giovanelli}, R., \& {Haynes}, M.~P. 1995, \aj, 109, 2415, \dodoi{10.1086/117460}

\bibitem[{{Chonis} {et~al.}(2016){Chonis}, {Hill}, {Lee}, {Tuttle}, {Vattiat}, {Drory}, {Indahl}, {Peterson}, \& {Ramsey}}]{LRS2}
{Chonis}, T.~S., {Hill}, G.~J., {Lee}, H., {et~al.} 2016, in Society of Photo-Optical Instrumentation Engineers (SPIE) Conference Series, Vol. 9908, Ground-based and Airborne Instrumentation for Astronomy VI, ed. C.~J. {Evans}, L.~{Simard}, \& H.~{Takami}, 99084C, \dodoi{10.1117/12.2232209}

\bibitem[{{Chung} {et~al.}(2009){Chung}, {van Gorkom}, {Kenney}, {Crowl}, \& {Vollmer}}]{Chung+2009}
{Chung}, A., {van Gorkom}, J.~H., {Kenney}, J. D.~P., {Crowl}, H., \& {Vollmer}, B. 2009, \aj, 138, 1741, \dodoi{10.1088/0004-6256/138/6/1741}

\bibitem[{{Cowie} \& {McKee}(1977)}]{Cowie+1977}
{Cowie}, L.~L., \& {McKee}, C.~F. 1977, \apj, 211, 135, \dodoi{10.1086/154911}

\bibitem[{{Davies} {et~al.}(2004){Davies}, {Minchin}, {Sabatini}, {van Driel}, {Baes}, {Boyce}, {de Blok}, {Disney}, {Evans}, {Kilborn}, {Lang}, {Linder}, {Roberts}, \& {Smith}}]{Davies+2004}
{Davies}, J., {Minchin}, R., {Sabatini}, S., {et~al.} 2004, \mnras, 349, 922, \dodoi{10.1111/j.1365-2966.2004.07568.x}

\bibitem[{{de Vaucouleurs} {et~al.}(1991){de Vaucouleurs}, {de Vaucouleurs}, {Corwin}, {Buta}, {Paturel}, \& {Fouqu{\'e}}}]{RC3}
{de Vaucouleurs}, G., {de Vaucouleurs}, A., {Corwin}, Jr., H.~G., {et~al.} 1991, {Third Reference Catalogue of Bright Galaxies. Volume I: Explanations and references. Volume II: Data for galaxies between 0$^{h}$ and 12$^{h}$. Volume III: Data for galaxies between 12$^{h}$ and 24$^{h}$.}

\bibitem[{{De Vis} {et~al.}(2019){De Vis}, {Jones}, {Viaene}, {Casasola}, {Clark}, {Baes}, {Bianchi}, {Cassara}, {Davies}, {De Looze}, {Galametz}, {Galliano}, {Lianou}, {Madden}, {Manilla-Robles}, {Mosenkov}, {Nersesian}, {Roychowdhury}, {Xilouris}, \& {Ysard}}]{DeVis+2019}
{De Vis}, P., {Jones}, A., {Viaene}, S., {et~al.} 2019, \aap, 623, A5, \dodoi{10.1051/0004-6361/201834444}

\bibitem[{{Dey} {et~al.}(2019){Dey}, {Schlegel}, {Lang}, {Blum}, {Burleigh}, {Fan}, {Findlay}, {Finkbeiner}, {Herrera}, {Juneau}, {Landriau}, {Levi}, {McGreer}, {Meisner}, {Myers}, {Moustakas}, {Nugent}, {Patej}, {Schlafly}, {Walker}, {Valdes}, {Weaver}, {Y{\`e}che}, {Zou}, {Zhou}, {Abareshi}, {Abbott}, {Abolfathi}, {Aguilera}, {Alam}, {Allen}, {Alvarez}, {Annis}, {Ansarinejad}, {Aubert}, {Beechert}, {Bell}, {BenZvi}, {Beutler}, {Bielby}, {Bolton}, {Brice{\~n}o}, {Buckley-Geer}, {Butler}, {Calamida}, {Carlberg}, {Carter}, {Casas}, {Castander}, {Choi}, {Comparat}, {Cukanovaite}, {Delubac}, {DeVries}, {Dey}, {Dhungana}, {Dickinson}, {Ding}, {Donaldson}, {Duan}, {Duckworth}, {Eftekharzadeh}, {Eisenstein}, {Etourneau}, {Fagrelius}, {Farihi}, {Fitzpatrick}, {Font-Ribera}, {Fulmer}, {G{\"a}nsicke}, {Gaztanaga}, {George}, {Gerdes}, {Gontcho}, {Gorgoni}, {Green}, {Guy}, {Harmer}, {Hernandez}, {Honscheid}, {Huang}, {James}, {Jannuzi}, {Jiang}, {Joyce}, {Karcher}, {Karkar}, {Kehoe}, {Kneib}, {Kueter-Young}, {Lan},
  {Lauer}, {Le Guillou}, {Le Van Suu}, {Lee}, {Lesser}, {Perreault Levasseur}, {Li}, {Mann}, {Marshall}, {Mart{\'\i}nez-V{\'a}zquez}, {Martini}, {du Mas des Bourboux}, {McManus}, {Meier}, {M{\'e}nard}, {Metcalfe}, {Mu{\~n}oz-Guti{\'e}rrez}, {Najita}, {Napier}, {Narayan}, {Newman}, {Nie}, {Nord}, {Norman}, {Olsen}, {Paat}, {Palanque-Delabrouille}, {Peng}, {Poppett}, {Poremba}, {Prakash}, {Rabinowitz}, {Raichoor}, {Rezaie}, {Robertson}, {Roe}, {Ross}, {Ross}, {Rudnick}, {Safonova}, {Saha}, {S{\'a}nchez}, {Savary}, {Schweiker}, {Scott}, {Seo}, {Shan}, {Silva}, {Slepian}, {Soto}, {Sprayberry}, {Staten}, {Stillman}, {Stupak}, {Summers}, {Sien Tie}, {Tirado}, {Vargas-Maga{\~n}a}, {Vivas}, {Wechsler}, {Williams}, {Yang}, {Yang}, {Yapici}, {Zaritsky}, {Zenteno}, {Zhang}, {Zhang}, {Zhou}, \& {Zhou}}]{Dey+2019}
{Dey}, A., {Schlegel}, D.~J., {Lang}, D., {et~al.} 2019, \aj, 157, 168, \dodoi{10.3847/1538-3881/ab089d}

\bibitem[{{Dolphin}(2016)}]{Dolphin2016}
{Dolphin}, A. 2016, {DOLPHOT: Stellar photometry}, Astrophysics Source Code Library, record ascl:1608.013.
\newblock \doeprint{1608.013}

\bibitem[{{Dolphin}(2000)}]{Dolphin2000}
{Dolphin}, A.~E. 2000, \pasp, 112, 1383, \dodoi{10.1086/316630}

\bibitem[{{Ferrarese} {et~al.}(2012){Ferrarese}, {C{\^o}t{\'e}}, {Cuillandre}, {Gwyn}, {Peng}, {MacArthur}, {Duc}, {Boselli}, {Mei}, {Erben}, {McConnachie}, {Durrell}, {Mihos}, {Jord{\'a}n}, {Lan{\c{c}}on}, {Puzia}, {Emsellem}, {Balogh}, {Blakeslee}, {van Waerbeke}, {Gavazzi}, {Vollmer}, {Kavelaars}, {Woods}, {Ball}, {Boissier}, {Courteau}, {Ferriere}, {Gavazzi}, {Hildebrandt}, {Hudelot}, {Huertas-Company}, {Liu}, {McLaughlin}, {Mellier}, {Milkeraitis}, {Schade}, {Balkowski}, {Bournaud}, {Carlberg}, {Chapman}, {Hoekstra}, {Peng}, {Sawicki}, {Simard}, {Taylor}, {Tully}, {van Driel}, {Wilson}, {Burdullis}, {Mahoney}, \& {Manset}}]{Ferrarese+2012}
{Ferrarese}, L., {C{\^o}t{\'e}}, P., {Cuillandre}, J.-C., {et~al.} 2012, \apjs, 200, 4, \dodoi{10.1088/0067-0049/200/1/4}

\bibitem[{{Fitzpatrick}(2004)}]{Fitzpatrick2004}
{Fitzpatrick}, E.~L. 2004, in Astronomical Society of the Pacific Conference Series, Vol. 309, Astrophysics of Dust, ed. A.~N. {Witt}, G.~C. {Clayton}, \& B.~T. {Draine}, 33, \dodoi{10.48550/arXiv.astro-ph/0401344}

\bibitem[{{Giovanelli} \& {Haynes}(1989)}]{Giovanelli+1989}
{Giovanelli}, R., \& {Haynes}, M.~P. 1989, \apjl, 346, L5, \dodoi{10.1086/185565}

\bibitem[{{Giovanelli} {et~al.}(2005){Giovanelli}, {Haynes}, {Kent}, {Perillat}, {Saintonge}, {Brosch}, {Catinella}, {Hoffman}, {Stierwalt}, {Spekkens}, {Lerner}, {Masters}, {Momjian}, {Rosenberg}, {Springob}, {Boselli}, {Charmandaris}, {Darling}, {Davies}, {Garcia Lambas}, {Gavazzi}, {Giovanardi}, {Hardy}, {Hunt}, {Iovino}, {Karachentsev}, {Karachentseva}, {Koopmann}, {Marinoni}, {Minchin}, {Muller}, {Putman}, {Pantoja}, {Salzer}, {Scodeggio}, {Skillman}, {Solanes}, {Valotto}, {van Driel}, \& {van Zee}}]{Giovanelli+2005}
{Giovanelli}, R., {Haynes}, M.~P., {Kent}, B.~R., {et~al.} 2005, \aj, 130, 2598, \dodoi{10.1086/497431}

\bibitem[{{G{\"o}ller} {et~al.}(2023){G{\"o}ller}, {Joshi}, {Rohr}, {Zinger}, \& {Pillepich}}]{Goller+2023}
{G{\"o}ller}, J., {Joshi}, G.~D., {Rohr}, E., {Zinger}, E., \& {Pillepich}, A. 2023, \mnras, 525, 3551, \dodoi{10.1093/mnras/stad2551}

\bibitem[{Green(2018)}]{dustmaps}
Green, G.~M. 2018, Journal of Open Source Software, 3, 695, \dodoi{10.21105/joss.00695}

\bibitem[{{Haynes} {et~al.}(2007){Haynes}, {Giovanelli}, \& {Kent}}]{Haynes+2007}
{Haynes}, M.~P., {Giovanelli}, R., \& {Kent}, B.~R. 2007, \apjl, 665, L19, \dodoi{10.1086/521188}

\bibitem[{{Haynes} {et~al.}(2011){Haynes}, {Giovanelli}, {Martin}, {Hess}, {Saintonge}, {Adams}, {Hallenbeck}, {Hoffman}, {Huang}, {Kent}, {Koopmann}, {Papastergis}, {Stierwalt}, {Balonek}, {Craig}, {Higdon}, {Kornreich}, {Miller}, {O'Donoghue}, {Olowin}, {Rosenberg}, {Spekkens}, {Troischt}, \& {Wilcots}}]{Haynes+2011}
{Haynes}, M.~P., {Giovanelli}, R., {Martin}, A.~M., {et~al.} 2011, \aj, 142, 170, \dodoi{10.1088/0004-6256/142/5/170}

\bibitem[{{Haynes} {et~al.}(2018){Haynes}, {Giovanelli}, {Kent}, {Adams}, {Balonek}, {Craig}, {Fertig}, {Finn}, {Giovanardi}, {Hallenbeck}, {Hess}, {Hoffman}, {Huang}, {Jones}, {Koopmann}, {Kornreich}, {Leisman}, {Miller}, {Moorman}, {O'Connor}, {O'Donoghue}, {Papastergis}, {Troischt}, {Stark}, \& {Xiao}}]{Haynes+2018}
{Haynes}, M.~P., {Giovanelli}, R., {Kent}, B.~R., {et~al.} 2018, \apj, 861, 49, \dodoi{10.3847/1538-4357/aac956}

\bibitem[{{Hill} {et~al.}(2021){Hill}, {Lee}, {MacQueen}, {Kelz}, {Drory}, {Vattiat}, {Good}, {Ramsey}, {Kriel}, {Peterson}, {DePoy}, {Gebhardt}, {Marshall}, {Tuttle}, {Bauer}, {Chonis}, {Fabricius}, {Froning}, {H{\"a}user}, {Indahl}, {Jahn}, {Landriau}, {Leck}, {Montesano}, {Prochaska}, {Snigula}, {Zeimann}, {Bryant}, {Damm}, {Fowler}, {Janowiecki}, {Martin}, {Mrozinski}, {Odewahn}, {Rostopchin}, {Shetrone}, {Spencer}, {Mentuch Cooper}, {Armandroff}, {Bender}, {Dalton}, {Hopp}, {Komatsu}, {Nicklas}, {Ramsey}, {Roth}, {Schneider}, {Sneden}, \& {Steinmetz}}]{HET2}
{Hill}, G.~J., {Lee}, H., {MacQueen}, P.~J., {et~al.} 2021, \aj, 162, 298, \dodoi{10.3847/1538-3881/ac2c02}

\bibitem[{{Huang} {et~al.}(2012){Huang}, {Haynes}, {Giovanelli}, \& {Brinchmann}}]{Huang+2012}
{Huang}, S., {Haynes}, M.~P., {Giovanelli}, R., \& {Brinchmann}, J. 2012, \apj, 756, 113, \dodoi{10.1088/0004-637X/756/2/113}

\bibitem[{{Hunter}(2007)}]{matplotlib}
{Hunter}, J.~D. 2007, Computing in Science and Engineering, 9, 90, \dodoi{10.1109/MCSE.2007.55}

\bibitem[{{Iglesias-P{\'a}ramo} {et~al.}(2006){Iglesias-P{\'a}ramo}, {Buat}, {Takeuchi}, {Xu}, {Boissier}, {Boselli}, {Burgarella}, {Madore}, {Gil de Paz}, {Bianchi}, {Barlow}, {Byun}, {Donas}, {Forster}, {Friedman}, {Heckman}, {Jelinski}, {Lee}, {Malina}, {Martin}, {Milliard}, {Morrissey}, {Neff}, {Rich}, {Schiminovich}, {Seibert}, {Siegmund}, {Small}, {Szalay}, {Welsh}, \& {Wyder}}]{Iglesias-Paramo+2006}
{Iglesias-P{\'a}ramo}, J., {Buat}, V., {Takeuchi}, T.~T., {et~al.} 2006, \apjs, 164, 38, \dodoi{10.1086/502628}

\bibitem[{{Jang} \& {Lee}(2017)}]{Jang+2017}
{Jang}, I.~S., \& {Lee}, M.~G. 2017, \apj, 835, 28, \dodoi{10.3847/1538-4357/835/1/28}

\bibitem[{{Jones} {et~al.}(2018){Jones}, {Papastergis}, {Pandya}, {Leisman}, {Romanowsky}, {Yung}, {Somerville}, \& {Adams}}]{Jones+2018}
{Jones}, M.~G., {Papastergis}, E., {Pandya}, V., {et~al.} 2018, \aap, 614, A21, \dodoi{10.1051/0004-6361/201732409}

\bibitem[{{Jones} {et~al.}(2022{\natexlab{a}}){Jones}, {Sand}, {Bellazzini}, {Spekkens}, {Cannon}, {Mutlu-Pakdil}, {Karunakaran}, {Beccari}, {Magrini}, {Cresci}, {Inoue}, {Fuson}, {Adams}, {Battaglia}, {Bennet}, {Crnojevi{\'c}}, {Caldwell}, {Guhathakurta}, {Haynes}, {Mu{\~n}oz}, {Seth}, {Strader}, {Toloba}, \& {Zaritsky}}]{Jones+2022}
{Jones}, M.~G., {Sand}, D.~J., {Bellazzini}, M., {et~al.} 2022{\natexlab{a}}, \apjl, 926, L15, \dodoi{10.3847/2041-8213/ac51dc}

\bibitem[{{Jones} {et~al.}(2022{\natexlab{b}}){Jones}, {Sand}, {Bellazzini}, {Spekkens}, {Karunakaran}, {Adams}, {Battaglia}, {Beccari}, {Bennet}, {Cannon}, {Cresci}, {Crnojevi{\'c}}, {Caldwell}, {Fuson}, {Guhathakurta}, {Haynes}, {Inoue}, {Magrini}, {Mu{\~n}oz}, {Mutlu-Pakdil}, {Seth}, {Strader}, {Toloba}, \& {Zaritsky}}]{Jones+2022b}
---. 2022{\natexlab{b}}, \apj, 935, 51, \dodoi{10.3847/1538-4357/ac7c6c}

\bibitem[{{Jones} {et~al.}(2023){Jones}, {Verdes-Montenegro}, {Moldon}, {Damas Segovia}, {Borthakur}, {Luna}, {Yun}, {del Olmo}, {Perea}, {Cannon}, {Lopez Gutierrez}, {Cluver}, {Garrido}, \& {Sanchez}}]{Jones+2023}
{Jones}, M.~G., {Verdes-Montenegro}, L., {Moldon}, J., {et~al.} 2023, \aap, 670, A21, \dodoi{10.1051/0004-6361/202244622}

\bibitem[{{Joye} \& {Mandel}(2003)}]{DS9}
{Joye}, W.~A., \& {Mandel}, E. 2003, in Astronomical Society of the Pacific Conference Series, Vol. 295, Astronomical Data Analysis Software and Systems XII, ed. H.~E. {Payne}, R.~I. {Jedrzejewski}, \& R.~N. {Hook}, 489

\bibitem[{{J{\'o}zsa} {et~al.}(2022){J{\'o}zsa}, {Jarrett}, {Cluver}, {Wong}, {Havenga}, {Yao}, {Marchetti}, {Taylor}, {Kamphuis}, {Maccagni}, {Ramaila}, {Serra}, {Smirnov}, {White}, {Kilborn}, {Holwerda}, {Hopkins}, {Brough}, {Pimbblet}, {Driver}, \& {Kuijken}}]{Jozsa+2022}
{J{\'o}zsa}, G.~I.~G., {Jarrett}, T.~H., {Cluver}, M.~E., {et~al.} 2022, \apj, 926, 167, \dodoi{10.3847/1538-4357/ac402b}

\bibitem[{{Kapferer} {et~al.}(2009){Kapferer}, {Sluka}, {Schindler}, {Ferrari}, \& {Ziegler}}]{Kapferer+2009}
{Kapferer}, W., {Sluka}, C., {Schindler}, S., {Ferrari}, C., \& {Ziegler}, B. 2009, \aap, 499, 87, \dodoi{10.1051/0004-6361/200811551}

\bibitem[{{Karunakaran} \& {Spekkens}(2024)}]{Karunakaran+2024}
{Karunakaran}, A., \& {Spekkens}, K. 2024, Research Notes of the American Astronomical Society, 8, 24, \dodoi{10.3847/2515-5172/ad1ee6}

\bibitem[{{Kenney} \& {Koopmann}(1999)}]{Kenney+1999}
{Kenney}, J. D.~P., \& {Koopmann}, R.~A. 1999, \aj, 117, 181, \dodoi{10.1086/300683}

\bibitem[{{Kenney} {et~al.}(2004){Kenney}, {van Gorkom}, \& {Vollmer}}]{Kenney+2004}
{Kenney}, J. D.~P., {van Gorkom}, J.~H., \& {Vollmer}, B. 2004, \aj, 127, 3361, \dodoi{10.1086/420805}

\bibitem[{{Kennicutt}(1998)}]{Kennicutt1998}
{Kennicutt}, Robert~C., J. 1998, \araa, 36, 189, \dodoi{10.1146/annurev.astro.36.1.189}

\bibitem[{{Kent} {et~al.}(2009){Kent}, {Spekkens}, {Giovanelli}, {Haynes}, {Momjian}, {Cort{\'e}s}, {Hardy}, \& {West}}]{Kent+2009}
{Kent}, B.~R., {Spekkens}, K., {Giovanelli}, R., {et~al.} 2009, \apj, 691, 1595, \dodoi{10.1088/0004-637X/691/2/1595}

\bibitem[{{Kent} {et~al.}(2007){Kent}, {Giovanelli}, {Haynes}, {Saintonge}, {Stierwalt}, {Balonek}, {Brosch}, {Catinella}, {Koopmann}, {Momjian}, \& {Spekkens}}]{Kent+2007}
{Kent}, B.~R., {Giovanelli}, R., {Haynes}, M.~P., {et~al.} 2007, \apjl, 665, L15, \dodoi{10.1086/521100}

\bibitem[{{Laher} {et~al.}(2012){Laher}, {Gorjian}, {Rebull}, {Masci}, {Fowler}, {Helou}, {Kulkarni}, \& {Law}}]{Laher+2012}
{Laher}, R.~R., {Gorjian}, V., {Rebull}, L.~M., {et~al.} 2012, \pasp, 124, 737, \dodoi{10.1086/666883}

\bibitem[{{Lee} {et~al.}(2022){Lee}, {Kimm}, {Blaizot}, {Katz}, {Lee}, {Sheen}, {Devriendt}, \& {Slyz}}]{Lee+2022}
{Lee}, J., {Kimm}, T., {Blaizot}, J., {et~al.} 2022, arXiv e-prints, arXiv:2201.01316.
\newblock \doarXiv{2201.01316}

\bibitem[{{Martin} {et~al.}(2005){Martin}, {Fanson}, {Schiminovich}, {Morrissey}, {Friedman}, {Barlow}, {Conrow}, {Grange}, {Jelinsky}, {Milliard}, {Siegmund}, {Bianchi}, {Byun}, {Donas}, {Forster}, {Heckman}, {Lee}, {Madore}, {Malina}, {Neff}, {Rich}, {Small}, {Surber}, {Szalay}, {Welsh}, \& {Wyder}}]{Martin+2005}
{Martin}, D.~C., {Fanson}, J., {Schiminovich}, D., {et~al.} 2005, \apjl, 619, L1, \dodoi{10.1086/426387}

\bibitem[{{McMullin} {et~al.}(2007){McMullin}, {Waters}, {Schiebel}, {Young}, \& {Golap}}]{CASA}
{McMullin}, J.~P., {Waters}, B., {Schiebel}, D., {Young}, W., \& {Golap}, K. 2007, in Astronomical Society of the Pacific Conference Series, Vol. 376, Astronomical Data Analysis Software and Systems XVI, ed. R.~A. {Shaw}, F.~{Hill}, \& D.~J. {Bell}, 127

\bibitem[{{Mei} {et~al.}(2007){Mei}, {Blakeslee}, {C{\^o}t{\'e}}, {Tonry}, {West}, {Ferrarese}, {Jord{\'a}n}, {Peng}, {Anthony}, \& {Merritt}}]{Mei+2007}
{Mei}, S., {Blakeslee}, J.~P., {C{\^o}t{\'e}}, P., {et~al.} 2007, \apj, 655, 144, \dodoi{10.1086/509598}

\bibitem[{{Millman} \& {Aivazis}(2011)}]{scipy2}
{Millman}, K.~J., \& {Aivazis}, M. 2011, Computing in Science and Engineering, 13, 9, \dodoi{10.1109/MCSE.2011.36}

\bibitem[{{Minchin} {et~al.}(2021){Minchin}, {Taylor}, \& {Deshev}}]{Minchin+2021}
{Minchin}, R., {Taylor}, R., \& {Deshev}, B. 2021, Research Notes of the American Astronomical Society, 5, 217, \dodoi{10.3847/2515-5172/ac29c7}

\bibitem[{{Minchin} {et~al.}(2005){Minchin}, {Davies}, {Disney}, {Boyce}, {Garcia}, {Jordan}, {Kilborn}, {Lang}, {Roberts}, {Sabatini}, \& {van Driel}}]{Minchin+2005}
{Minchin}, R., {Davies}, J., {Disney}, M., {et~al.} 2005, \apjl, 622, L21, \dodoi{10.1086/429538}

\bibitem[{{Minchin} {et~al.}(2019){Minchin}, {Taylor}, {K{\"o}ppen}, {Davies}, {van Driel}, \& {Keenan}}]{Minchin+2019}
{Minchin}, R.~F., {Taylor}, R., {K{\"o}ppen}, J., {et~al.} 2019, \aj, 158, 121, \dodoi{10.3847/1538-3881/ab303e}

\bibitem[{{Morrissey} {et~al.}(2007){Morrissey}, {Conrow}, {Barlow}, {Small}, {Seibert}, {Wyder}, {Budav{\'a}ri}, {Arnouts}, {Friedman}, {Forster}, {Martin}, {Neff}, {Schiminovich}, {Bianchi}, {Donas}, {Heckman}, {Lee}, {Madore}, {Milliard}, {Rich}, {Szalay}, {Welsh}, \& {Yi}}]{Morrissey+2007}
{Morrissey}, P., {Conrow}, T., {Barlow}, T.~A., {et~al.} 2007, \apjs, 173, 682, \dodoi{10.1086/520512}

\bibitem[{{Mosenkov} {et~al.}(2019){Mosenkov}, {Baes}, {Bianchi}, {Casasola}, {Cassar{\`a}}, {Clark}, {Davies}, {De Looze}, {De Vis}, {Fritz}, {Galametz}, {Galliano}, {Jones}, {Lianou}, {Madden}, {Nersesian}, {Smith}, {Tr{\v{c}}ka}, {Verstocken}, {Viaene}, {Vika}, \& {Xilouris}}]{Mosenkov+2019}
{Mosenkov}, A.~V., {Baes}, M., {Bianchi}, S., {et~al.} 2019, \aap, 622, A132, \dodoi{10.1051/0004-6361/201833932}

\bibitem[{{Oliphant}(2007)}]{scipy1}
{Oliphant}, T.~E. 2007, Computing in Science and Engineering, 9, 10, \dodoi{10.1109/MCSE.2007.58}

\bibitem[{{O'Neil} {et~al.}(2024){O'Neil}, {van Driel}, \& {Schneider}}]{O'Neil+2024}
{O'Neil}, K., {van Driel}, W., \& {Schneider}, S. 2024, in American Astronomical Society Meeting Abstracts, Vol.~56, American Astronomical Society Meeting Abstracts, 404.13

\bibitem[{{Osterbrock}(1989)}]{Osterbrock}
{Osterbrock}, D.~E. 1989, {Astrophysics of gaseous nebulae and active galactic nuclei}

\bibitem[{pandas~development team(2020)}]{pandas2}
pandas~development team, T. 2020, pandas-dev/pandas: Pandas, latest,  Zenodo, \dodoi{10.5281/zenodo.3509134}

\bibitem[{{Pettini} \& {Pagel}(2004)}]{Pettini+2004}
{Pettini}, M., \& {Pagel}, B. E.~J. 2004, \mnras, 348, L59, \dodoi{10.1111/j.1365-2966.2004.07591.x}

\bibitem[{{Ramsey} {et~al.}(1998){Ramsey}, {Adams}, {Barnes}, {Booth}, {Cornell}, {Fowler}, {Gaffney}, {Glaspey}, {Good}, {Hill}, {Kelton}, {Krabbendam}, {Long}, {MacQueen}, {Ray}, {Ricklefs}, {Sage}, {Sebring}, {Spiesman}, \& {Steiner}}]{HET1}
{Ramsey}, L.~W., {Adams}, M.~T., {Barnes}, T.~G., {et~al.} 1998, in Society of Photo-Optical Instrumentation Engineers (SPIE) Conference Series, Vol. 3352, Advanced Technology Optical/IR Telescopes VI, ed. L.~M. {Stepp}, 34--42, \dodoi{10.1117/12.319287}

\bibitem[{{Robitaille} {et~al.}(2020){Robitaille}, {Deil}, \& {Ginsburg}}]{reproject}
{Robitaille}, T., {Deil}, C., \& {Ginsburg}, A. 2020, {reproject: Python-based astronomical image reprojection}.
\newblock \doeprint{2011.023}

\bibitem[{{Sand} {et~al.}(2015){Sand}, {Crnojevi{\'c}}, {Bennet}, {Willman}, {Hargis}, {Strader}, {Olszewski}, {Tollerud}, {Simon}, {Caldwell}, {Guhathakurta}, {James}, {Koposov}, {McLeod}, {Morrell}, {Peacock}, {Salinas}, {Seth}, {Stark}, \& {Toloba}}]{Sand+2015}
{Sand}, D.~J., {Crnojevi{\'c}}, D., {Bennet}, P., {et~al.} 2015, \apj, 806, 95, \dodoi{10.1088/0004-637X/806/1/95}

\bibitem[{{Sand} {et~al.}(2017){Sand}, {Seth}, {Crnojevi{\'c}}, {Spekkens}, {Strader}, {Adams}, {Caldwell}, {Guhathakurta}, {Kenney}, {Randall}, {Simon}, {Toloba}, \& {Willman}}]{Sand+2017}
{Sand}, D.~J., {Seth}, A.~C., {Crnojevi{\'c}}, D., {et~al.} 2017, \apj, 843, 134, \dodoi{10.3847/1538-4357/aa7557}

\bibitem[{{Schlafly} \& {Finkbeiner}(2011)}]{Schlafly+2011}
{Schlafly}, E.~F., \& {Finkbeiner}, D.~P. 2011, \apj, 737, 103, \dodoi{10.1088/0004-637X/737/2/103}

\bibitem[{{Schlegel} {et~al.}(1998){Schlegel}, {Finkbeiner}, \& {Davis}}]{Schlegel+1998}
{Schlegel}, D.~J., {Finkbeiner}, D.~P., \& {Davis}, M. 1998, \apj, 500, 525, \dodoi{10.1086/305772}

\bibitem[{{Serra} {et~al.}(2014){Serra}, {Westmeier}, {Giese}, {Jurek}, {Fl{\"o}er}, {Popping}, {Winkel}, {van der Hulst}, {Meyer}, {Koribalski}, {Staveley-Smith}, \& {Courtois}}]{SoFiA}
{Serra}, P., {Westmeier}, T., {Giese}, N., {et~al.} 2014, {SoFiA: Source Finding Application}, Astrophysics Source Code Library.
\newblock \doeprint{1412.001}

\bibitem[{{Serra} {et~al.}(2015){Serra}, {Westmeier}, {Giese}, {Jurek}, {Fl{\"o}er}, {Popping}, {Winkel}, {van der Hulst}, {Meyer}, {Koribalski}, {Staveley-Smith}, \& {Courtois}}]{Serra+2015}
---. 2015, \mnras, 448, 1922, \dodoi{10.1093/mnras/stv079}

\bibitem[{{Sirianni} {et~al.}(2005){Sirianni}, {Jee}, {Ben{\'\i}tez}, {Blakeslee}, {Martel}, {Meurer}, {Clampin}, {De Marchi}, {Ford}, {Gilliland}, {Hartig}, {Illingworth}, {Mack}, \& {McCann}}]{Sirianni+2005}
{Sirianni}, M., {Jee}, M.~J., {Ben{\'\i}tez}, N., {et~al.} 2005, \pasp, 117, 1049, \dodoi{10.1086/444553}

\bibitem[{{Sorgho} {et~al.}(2017){Sorgho}, {Hess}, {Carignan}, \& {Oosterloo}}]{Sorgho+2017}
{Sorgho}, A., {Hess}, K., {Carignan}, C., \& {Oosterloo}, T.~A. 2017, \mnras, 464, 530, \dodoi{10.1093/mnras/stw2341}

\bibitem[{{Taylor} {et~al.}(2011){Taylor}, {Hopkins}, {Baldry}, {Brown}, {Driver}, {Kelvin}, {Hill}, {Robotham}, {Bland-Hawthorn}, {Jones}, {Sharp}, {Thomas}, {Liske}, {Loveday}, {Norberg}, {Peacock}, {Bamford}, {Brough}, {Colless}, {Cameron}, {Conselice}, {Croom}, {Frenk}, {Gunawardhana}, {Kuijken}, {Nichol}, {Parkinson}, {Phillipps}, {Pimbblet}, {Popescu}, {Prescott}, {Sutherland}, {Tuffs}, {van Kampen}, \& {Wijesinghe}}]{Taylor+2011}
{Taylor}, E.~N., {Hopkins}, A.~M., {Baldry}, I.~K., {et~al.} 2011, \mnras, 418, 1587, \dodoi{10.1111/j.1365-2966.2011.19536.x}

\bibitem[{{Taylor} {et~al.}(2012){Taylor}, {Davies}, {Auld}, \& {Minchin}}]{Taylor+2012}
{Taylor}, R., {Davies}, J.~I., {Auld}, R., \& {Minchin}, R.~F. 2012, \mnras, 423, 787, \dodoi{10.1111/j.1365-2966.2012.20914.x}

\bibitem[{{Tonnesen} \& {Bryan}(2012)}]{Tonnesen+2012}
{Tonnesen}, S., \& {Bryan}, G.~L. 2012, \mnras, 422, 1609, \dodoi{10.1111/j.1365-2966.2012.20737.x}

\bibitem[{{Tonnesen} \& {Bryan}(2021)}]{Tonnesen+2021}
---. 2021, \apj, 911, 68, \dodoi{10.3847/1538-4357/abe7e2}

\bibitem[{{van der Walt} {et~al.}(2011){van der Walt}, {Colbert}, \& {Varoquaux}}]{numpy}
{van der Walt}, S., {Colbert}, S.~C., \& {Varoquaux}, G. 2011, Computing in Science and Engineering, 13, 22, \dodoi{10.1109/MCSE.2011.37}

\bibitem[{{Verdes-Montenegro} {et~al.}(2001){Verdes-Montenegro}, {Yun}, {Williams}, {Huchtmeier}, {Del Olmo}, \& {Perea}}]{Verdes-Montenegro+2001}
{Verdes-Montenegro}, L., {Yun}, M.~S., {Williams}, B.~A., {et~al.} 2001, \aap, 377, 812, \dodoi{10.1051/0004-6361:20011127}

\bibitem[{{Vollmer} {et~al.}(2004){Vollmer}, {Beck}, {Kenney}, \& {van Gorkom}}]{Vollmer+2004}
{Vollmer}, B., {Beck}, R., {Kenney}, J. D.~P., \& {van Gorkom}, J.~H. 2004, \aj, 127, 3375, \dodoi{10.1086/420802}

\bibitem[{{Vollmer} {et~al.}(2001){Vollmer}, {Cayatte}, {Balkowski}, \& {Duschl}}]{Vollmer+2001}
{Vollmer}, B., {Cayatte}, V., {Balkowski}, C., \& {Duschl}, W.~J. 2001, \apj, 561, 708, \dodoi{10.1086/323368}

\bibitem[{{Vollmer} {et~al.}(2006){Vollmer}, {Soida}, {Otmianowska-Mazur}, {Kenney}, {van Gorkom}, \& {Beck}}]{Vollmer+2006}
{Vollmer}, B., {Soida}, M., {Otmianowska-Mazur}, K., {et~al.} 2006, \aap, 453, 883, \dodoi{10.1051/0004-6361:20064954}

\bibitem[{{W}es {M}c{K}inney(2010)}]{pandas1}
{W}es {M}c{K}inney. 2010, in {P}roceedings of the 9th {P}ython in {S}cience {C}onference, ed. {S}t\'efan van~der {W}alt \& {J}arrod {M}illman, 56 -- 61, \dodoi{10.25080/Majora-92bf1922-00a}

\bibitem[{{Wong} {et~al.}(2021){Wong}, {Stevens}, {For}, {Westmeier}, {Dixon}, {Oh}, {J{\'o}zsa}, {Reynolds}, {Lee-Waddell}, {Rom{\'a}n}, {Verdes-Montenegro}, {Courtois}, {Pomar{\`e}de}, {Murugeshan}, {Whiting}, {Bekki}, {Bigiel}, {Bosma}, {Catinella}, {D{\'e}nes}, {Elagali}, {Holwerda}, {Kamphuis}, {Kilborn}, {Kleiner}, {Koribalski}, {Lelli}, {Madrid}, {McQuinn}, {Popping}, {Rhee}, {Roychowdhury}, {Scott}, {Sengupta}, {Spekkens}, {Staveley-Smith}, \& {Wakker}}]{Wong+2021}
{Wong}, O.~I., {Stevens}, A.~R.~H., {For}, B.~Q., {et~al.} 2021, \mnras, 507, 2905, \dodoi{10.1093/mnras/stab2262}

\bibitem[{{Wyder} {et~al.}(2007){Wyder}, {Martin}, {Schiminovich}, {Seibert}, {Budav{\'a}ri}, {Treyer}, {Barlow}, {Forster}, {Friedman}, {Morrissey}, {Neff}, {Small}, {Bianchi}, {Donas}, {Heckman}, {Lee}, {Madore}, {Milliard}, {Rich}, {Szalay}, {Welsh}, \& {Yi}}]{Wyder+2007}
{Wyder}, T.~K., {Martin}, D.~C., {Schiminovich}, D., {et~al.} 2007, \apjs, 173, 293, \dodoi{10.1086/521402}

\bibitem[{{Zhou} {et~al.}(2023){Zhou}, {Zhu}, {Yang}, {Yu}, {Yuan}, {Jiang}, \& {Xi}}]{Zhou+2023}
{Zhou}, R., {Zhu}, M., {Yang}, Y., {et~al.} 2023, \apj, 952, 130, \dodoi{10.3847/1538-4357/acdcf5}

\bibitem[{{Zibetti} {et~al.}(2009){Zibetti}, {Charlot}, \& {Rix}}]{Zibetti+2009}
{Zibetti}, S., {Charlot}, S., \& {Rix}, H.-W. 2009, \mnras, 400, 1181, \dodoi{10.1111/j.1365-2966.2009.15528.x}

\end{thebibliography}
\bibliographystyle{aasjournal}



\end{document}